\documentclass[twocolumn]{aa}
\usepackage{graphicx}
\usepackage{txfonts}
\usepackage{aalongtable}
\begin{document}
   \title{L-band (3.5 $\mu$m) IR-excess in massive star formation}

   \subtitle{I. 30 Doradus}

   \author{M. Maercker
          \inst{1,2}
          \and
          M. G. Burton\inst{2}}

   \offprints{M. Maercker}

   \institute{Stockholm Observatory, AlbaNova University Center,\\
	     106 91 Stockholm, Sweden\\
             \email{maercker@astro.su.se}
         \and
             School of Physics, University of New South Wales,\\
              Sydney, NSW 2052, Australia\\
              \email{mgb@phys.unsw.edu.au}}

   \date{Received January 18, 2005; accepted April 18, 2005}

   \abstract{L-band data of 30 Doradus at 3.5 $\mu$m taken with SPIREX (South Pole
Infrared Explorer) is presented. The photometry was combined with 2MASS JHK data 
at 1.25 -2.2 $\mu$m. Colour-colour and colour-magnitude diagrams are constructed and
used to determine the sources with infrared excess. These are interpreted as 
circumstellar disks, and enable the fraction of sources with disks (the cluster 
disk fraction or CDF) to be determined. We find that  $\sim42\%$ of the sources
detected at L-band in 30 Doradus have an IR-excess.\thanks{Table 6 is also available
in electronic form at the CDS via anonymous ftp to cdsarc.u-strasbg.fr (130.79.128.5)
or via http://cdsweb.u-strasbg.fr/cgi-bin/qcat?J/A+A/} 
   
   }

   \maketitle
%
\section{Introduction}
\subsection{IR-excess as a measure of circumstellar disks}
During star formation, young stellar objects (YSOs) are associated with the 
circumstellar material surrounding them, making them bright at infrared wavelengths 
as the dust absorbs and re-emits radiation from the central star. As the star evolves
towards the main sequence, the distribution of the surrounding material also changes,
as is evident through the changing spectral energy distribution (SED) of the source. 
SEDs can therefore be used as an indicator of the evolutionary state 
of a YSO. In particular, an excess in IR radiation above that of a
blackbody for class II/III YSOs can be explained by models of circumstellar disks
around protostars (eg. Lada \& Adams \cite{ladaadams}). In IR colour-colour diagrams, 
such stars lie outside the region defined by the reddening band to the main sequence.
Colour-colour diagrams can therefore be used to identify the stars with 
circumstellar disks in star forming regions, and to estimate what fraction of the
population this represents.

\subsection{L-band over K-band}
The excess emission is measurable in the K-band at 2.2 $\mu$m, making 
infrared excess detectable using JHK (1-2.2 $\mu$m) observations. However, this 
radiation may be caused by a circumstellar disk, the protostellar envelope or 
emission nebulosity from HII regions and therefore JHK band observations alone 
may not be sufficient to determine the nature of the IR excess. L-band observations 
(3.5 $\mu$m) prove to be the ideal wavelength for detecting circumstellar disks
(eg. Kenyon \& Hartmann \cite{kenyonhartmann}). 
Compared to JHK colour-colour diagrams, the stars with IR excess are much more clearly 
separated in JHKL colour-colour diagrams. Additionally, the continuum emission 
from the stellar photosphere is generally bright enough in L-band to allow the 
determination of the number of stars with IR excess compared to the total 
number of stars in a given region (i.e. the cluster disk fraction (CDF)). It is
relatively easier to obtain L-band data with the same sensitivities and spatial 
resolution as JHK band observations than it is for mid-IR data, where the signature from 
disk emission is even more pronounced (eg. Rathborne \cite{rathborne}).

\subsection{30 Doradus}
\label{30dor}
30 Doradus in the Large Magellanic Cloud is the most luminous giant HII region in
the Local Group of Galaxies (Kennicutt~\cite{kennicutt}) and is located at 
a distance of $\sim$55 kpc (Vermeij et al.~\cite{vermeijetal}). Star formation within the 
region was first identified by Hyland et al. (\cite{hylandetal}). Walborn and Blades 
(\cite{walbornblades}) identify four phases comprising an age sequence for massive OB cluster 
evolution (Walborn~\cite{walborna}, \cite{walbornb}) in five spatial and/or temporal structures. 
The region shows recent and ongoing star formation at discrete epochs, with a central cluster
of massive stars surrounded by extended nebulosity. The stellar population
consists of multiple generations, with pre-main sequence stars, early type main 
sequence stars and evolved blue and red supergiants (Walborn \& Blades~\cite{walbornblades}). 
Ages range from young stars less than 1 Myr old, to the giant population with ages up to 25 Myr. 
Two supernova remnant (SNR) candidates have also been identified (Lazendic et al.
\cite{lazendicetal}). Using deep broadband I and V WFPC2 images from the HST, 
Sirianni et al. (\cite{sirianni}) derived the IMF for the central cluster R136 in
30 Doradus down to ~1.35 M$_{\sun}$, suggesting there are relatively fewer lower-mass 
stars than in the Galactic IMF (though this may also arise from not fully correcting
for extinction from the source). IR and radio observations show molecular gas and warm 
dust concentrated in an arc to the north and west of the central cluster (Rubio et
al \cite{rubio}). They found 84 IR sources linked with the nebular microstructures and 
early O stars in dense nebular knots. 6 of these could be matched with sources found in
this paper (Table~\ref{photres}). Identification of early O-type stars was taken 
as evidence for star formation (Walborn \& Blades \cite{walbornblades}). These knots 
have been revealed to be compact multiple systems which are each similar to star forming 
regions in the Galaxy such as the Trapezium system in the Orion cluster (Brandner et al. 
\cite{brandner}). In their paper Brandner et al. use near-IR colour-colour diagrams
obtained from NICMOS data to identify pre-main sequence stars by their intrinsic 
IR-excess. It is suggested that the spatial distribution of these stars indicate the 
birth of an OB association. Present day star formation coincides with the dense regions
of molecular gas (Johansson et al.~\cite{johanssonetal}) to the north and west of R136 
and with their interfaces with the cavity created by the central cluster
(Brandner et al.~\cite{brandner}).
It is the nearest and therefore most highly resolved starburst region outside the Galaxy, 
thus making it a suitable source to study extragalactic star formation.

This paper presents L-band photometry of the 30 Doradus region taken with the SPIREX
telescope at the South-Pole. IR-excess is detected by combining the L-band data with 
JHK-data from 2MASS. Colour-colour and colour-magnitude diagrams are presented,
with a discussion on the implications for massive star formation and the
evolution of circumstellar disks. Section~\ref{observations} describes the  
observational data. Section~\ref{results} presents the results of the photometry
on the L-band images. Section~\ref{analysis} analyses the results and 
Section~\ref{discussion} provides an interpretation.

\section{Observations}
\label{observations}
\subsection{L-band data from SPIREX}
The L-band data was taken using the 60 cm South Pole InfraRed Explorer (SPIREX) 
(Hereld~\cite{hereld}; Burton et al.~\cite{burtonetal}) at the Amundsen-Scott South-Pole 
station in the Antarctica in 1998 by the winter-over scientist Charlie Kaminsky. The 
telescope was equipped with an Aladdin 1024x1024 InSb Abu detector with filters sensitive
in the range from 2.4 - 5 $\mu$m. The camera provided a $10'$ field of view with a 
pixel size of 0.6\arcsec. The diffraction limit of 1.4\arcsec, combined with seeing
and tracking error, over the unguided observations, resulted in a typical resolution of 
2.6\arcsec. Observations were taken through the L-band filter 
(${\lambda}_{central}=3.514~\mu$m, $\Delta\lambda=0.618~\mu$m) by taking one set of sky 
frames followed by two sets of object frames and another set of sky frames. Each set 
consisted of five averaged frames offset by approximately $30\arcsec$ from the previous 
frame. This allowed for the easy removal of image artefacts from the array and sky 
emission and of stars in the sky frames. The total on-source integration time was 9.25 
hours. Reductions were done automatically by the RIT SPIREX/Abu 
pipeline\footnote{http://pipe.cis.rit.edu}. This image was archival data from SPIREX,
but unfortunately had not been flux-calibrated, requiring us to determine the calibration
through seperate observations (see Sec.~\ref{calcasp}).

\subsection{JHK-band data from 2MASS}
The observations from SPIREX were complemented with the 2MASS point source 
catalogue (PSC) (Cutri et al.,~\cite{cutrietal}) and Atlas Images
\footnote{Available at http://www.ipac.caltech.edu/applications/2MASS/IM/}. 
The 2MASS telescopes scanned the sky in both 
hemispheres in three near infrared filters (J, H and K; 1.25, 1.65 and 2.2 
$\mu$m respectively) and detected point sources in each band which were 
brighter than about 1mJy using a pixel size of 2.0\arcsec. Two 1.3 metre telescopes
were used, located at Mt Hopkins in Arizona and CTIO in Chile. Each used a camera 
capable of observing in the bands simultaneously with the 256x256 arrays of HgCdTe 
detectors. The imaging was done while the highly automated telescopes 
scanned over the sky in declination at a rate of $\sim1\arcmin$ per second. The images 
consist of six pointings on the sky with a total integration time of 7.8 
seconds (Kleinmann et al.~\cite{kleinmannetal}). The Atlas Images from the 2MASS catalogue 
were used to derive K-band magnitudes for sources seen in the L-band that could not be 
matched with the sources in the PSC.

\subsection{Calibration data from CASPIR}
\label{calcasp}
In order to calibrate the SPIREX image, observations were carried out in early April 
2004 using the Australian National University (ANU) 2.3m telescope at Siding Spring 
Observatory, equipped with CASPIR (Cryogenic Array Spectrometer/Imager)
(McGregor~\cite{mcgregor}). CASPIR is a cryogenic re-imaging camera with pixel scales 
of $0.5\arcsec$/pixel and $0.25\arcsec$/pixel and 2 different readout methods. 
The detector is a hybrid device with 256x256 pixels, capable of direct imaging and 
spectroscopy between 1-5 $\mu$m. To avoid saturation from the sky in the L-band, a 
narrow band filter (${\lambda}_{central}=3.592 ~\mu$m, $\Delta\lambda=0.078~\mu$m) had to 
be used as well as the smaller pixel scale, resulting in a $60\arcsec$ field of view. The 
standard star used is listed in Table~\ref{standards}. The stars in 30 Doradus that were 
used to calibrate the remaining stars in the L-band images are listed in Table~\ref{cali}.
These are bright and isolated stars in the SPIREX image. Comparison of the relative photometry
of the five stars observed both with CASPIR and SPIREX indicates that individual errors of 
$\pm0.1$ to $\pm0.2$ magnitudes are made, and taking the weighted average this leads to a zero 
point error of 0.04 mags. This has been included in all subsequent error calulcation.

\begin{table}
\caption{Standard star used to calibrate the CASPIR images.}
\label{standards}
\centering
\begin{tabular}{c c c c}
\hline\hline
name & RA (J2000) & DEC(J2000) & $m_L$\\
 & (h m s) & (d m s) & Mag\\
\hline
BS2015 & 05 44 46.5 & -65 44 08.0 & 3.711\\
\hline
\end{tabular}
\end{table}

\begin{table}
\caption{Stars in 30 Doradus used for calibration. Bright, isolated stars were 
chosen from the SPIREX image and used to calibrate the remaining stars in the 
image. Their L-band magnitudes and errors were determined from the standard star.}
\label{cali}
\centering
\begin{tabular}{c c c c}
\hline\hline
id & RA (J2000) & DEC(J2000) & $m_{L}$\\
 & (h m s) & (d m s) & Mag \\
\hline
11 & 05 38 06.6 & -69 03 45.0 & 9.3 $\pm$ 0.2\\
14 & 05 38 09.6 & -69 06 21.2 & 8.7 $\pm$ 0.1\\
26 & 05 38 16.7 & -69 04 14.2 & 8.3 $\pm$ 0.1\\
28 & 05 38 17.0 & -69 04 00.8 & 8.7 $\pm$ 0.2\\
29 & 05 38 17.6 & -69 04 12.0 & 9.6 $\pm$ 0.1\\
\hline
\end{tabular}
\end{table}

\section{Results}
\label{results}

   \begin{figure*}
   \centering
   \includegraphics[width=12cm]{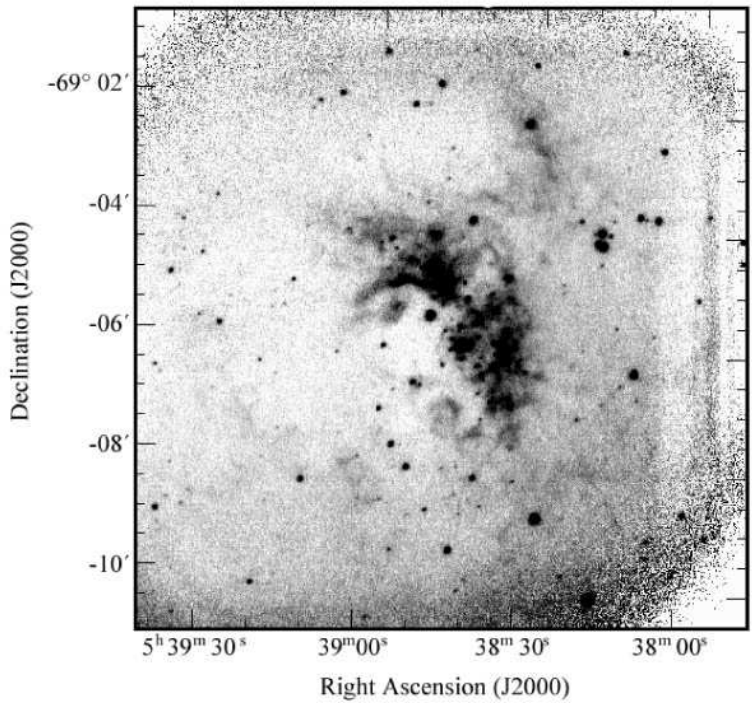}
   \caption{SPIREX L-band (3.5$\mu m$) image of 30 Doradus. Total integration time 9.25 hours;
effective resolution 2.6$\arcsec$; pixel scale 0.6$\arcsec$; 90\% completeness limit at 13.5 mag;
faintest star detected 14.5 mag.}
              \label{30dorL}
    \end{figure*}

\begin{figure*}
\centering
\includegraphics[width=12cm]{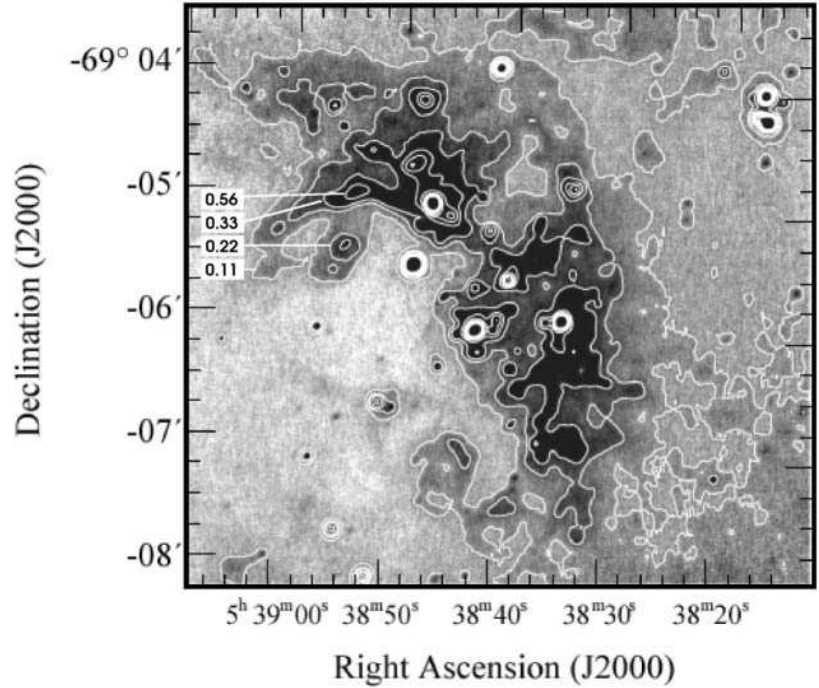}
\caption{Enlargements of the central region of 30 Doradus with contours showing the
nebulosity. Contour levels are 0.11, 0.22, 0.33, 0.56, 0.75, 0.97, 1.19, 1.42  and 
1.64 mJy/arcsec$^2$. The first four levels are labeled in the image. The image shows the nebulous arcs 
and denser knots in the central region. These denser regions coincide with the distribution of 
molecular gas (Johansson et al.~\protect\cite{johanssonetal}) and are believed to be regions of present day 
star formation.}
\label{30dorcont}
\end{figure*}

\begin{figure*}
\centering
\includegraphics[width=12cm]{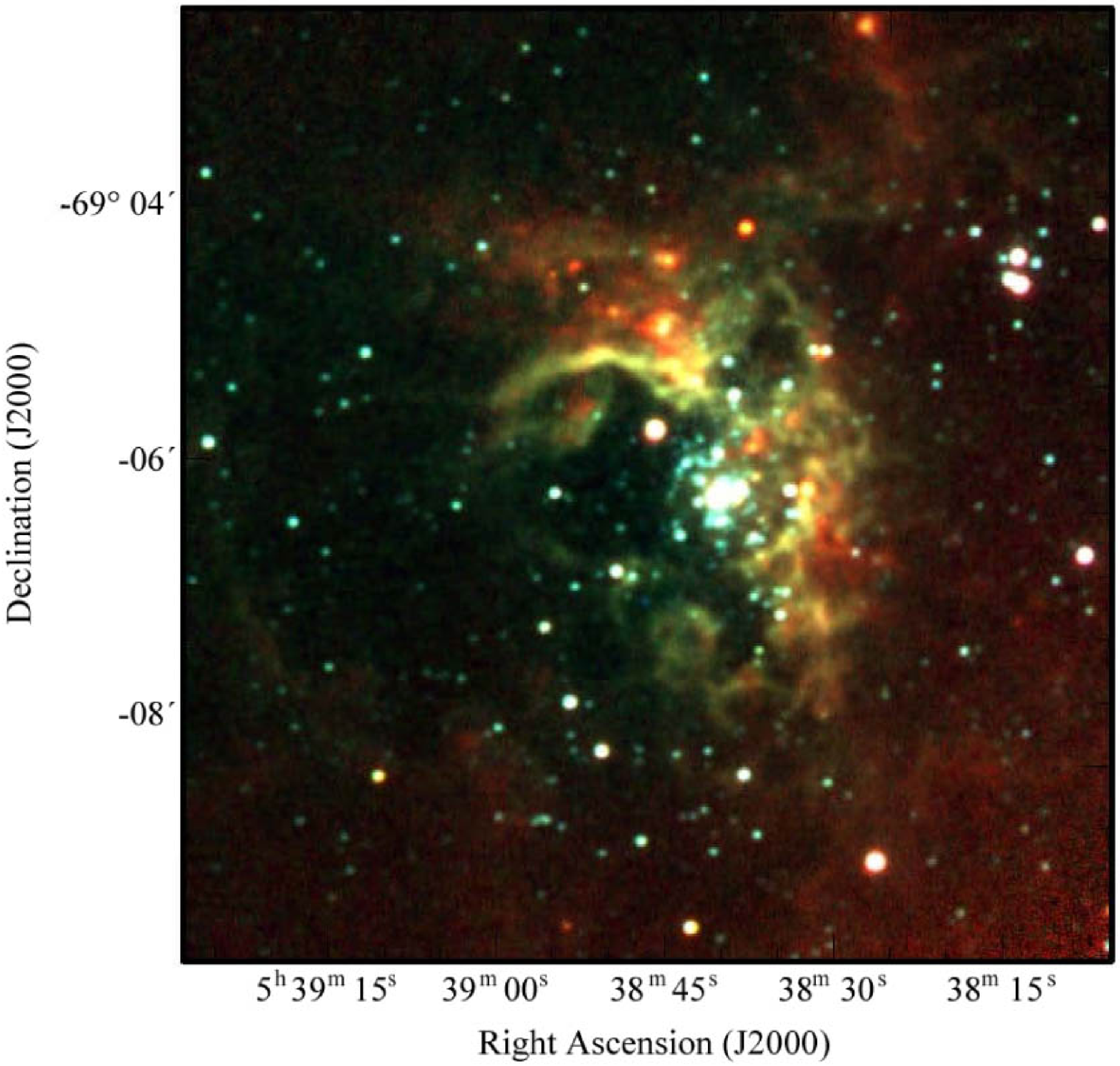}
\caption{HKL (Blue=H, Green=K, Red=L) composite colour image of 30 Doradus created using the 
2MASS and SPIREX images. Regions bright in the L-band ($3.5\mu$m) can be seen to the north-east 
and west of the central cluster, indicating the presence of young stellar objects.}
\label{colour}
\end{figure*}

\subsection{Photometry}
The SPIREX L-band image of 30 Doradus is shown in Fig.~\ref{30dorL}. Figure~\ref{30dorcont} 
shows an enlargement of the central region overlaid with contours showing the nebulosity. 
A coordinate frame was fit to the SPIREX images using the \emph{koords} program in the 
\emph{Karma} package\footnote{http://www.atnf.csiro.au/karma/} 
and the 2MASS atlas images for the reference frame. Photometry was 
undertaken using the \emph{IRAF/daophot} package\footnote{http://iraf.noao.edu/}. The 
resulting fluxes were calibrated using the stars measured with CASPIR, to determine
the zero point correction. Error estimation was done using the \emph{addstar} 
package by adding 220 artificial stars and running the same analysis as on the 
original image. This resulted in a 90\% completeness limit of 13.5 mag in L-band.
In order to include stars from the 2MASS images possibly missed by 
the automated detection process used to create the PSC, the same analysis was run
on the 2MASS K-band images. This resulted in 7 additional matches for sources in
the L-band image that were not matched with the PSC.

Matches between the SPIREX and 2MASS images were confirmed visually by 
overlaying and blinking the images. For 24 stars no match for the L-band
stars could be found with the JHK data. Table~\ref{phot} summarizes the statistics
of the detections in the various bands. The results of the photometry including errors
are given in Table~\ref{photres}. Table~\ref{phot} also lists the statistics in the 
various bands for detections brighter than the 90\% L-band completeness limit.

Figure~\ref{colour} shows a HKL colour image of the central region, showing the bright
features seen in the L-band, indicating regions of massive star formation.

\subsection{Sensitivity}
\label{sensitivity}
The detection threshold was assumed to be three times the background variation
${\sigma}_{sky}$, the typical standard deviation of the background in a sky
region near each source. This provides a limiting magnitude of m$_{limit} = 13.5$. 
This corresponds to the 90\% completeness limit determined by adding artificial
stars. However, the background is variable over the image, particularly on account of
nebulosity and source confusion in the more crowded regions. The faintest source
detected has an L-band magnitude of 14.5, corresponding to a 78\% completeness limit. This 
sensitivity has only recently been bettered by a ground-based telescope; Stolte et al. 
(\cite{stolteetal}) achieved a limiting magnitude of 15 at L-band using the 8m VLT. This 
illustrates the improved sensitivity as a result of the low thermal background in Antarctica 
(100-300 $mJy/arcsec^2$ at 3.5$\mu m$; Phillips et al.~\cite{phillipsetal}).

\subsection{Foreground contamination}
\label{foreground}
Off-source comparison images were not available for the SPIREX images. An 
estimation of the foreground contamination therefore had to be made using the
(J-K) colours of the stars. If the visual extinction $A_{V}$ to the source is 
known, the (J-K) colour excess due to interstellar reddening between the Earth and 
the source can be calculated. Assuming that the majority of all source stars are 
embedded and therefore additionally reddened, any star bluer than this (J-K) 
colour is likely to be a foreground star. Adopting an extinction parameter of 
$R_{V}=\frac{A_{V}}{E(B-V)}$ of $\sim3.4$ (Selman et al. \cite{selmanetal}) 
and the (B-V) and (B-V)$_{0}$ colours for sources in 30 Doradus (Walborn \& 
Blades \cite{walbornblades}) gives a visual extinction of $\sim1.4$ which 
corresponds to a limiting (J-K) colour of $\sim0.25$. 

An additional estimation of the (J-K) colour of likely foreground stars was done using 
only the 2MASS data. 8 off-source fields with $3\arcmin$ radii were extracted from the 
PSC and compared to a source field centred on 30 Doradus with a $7\arcmin$ radius to 
encompass the entire area of the SPIREX image. The off-source fields were combined and 
the data was then plotted in JHK colour-colour diagrams and the number of IR-excess 
sources was determined. To minimise scatter in the colour-colour diagrams, only sources 
that lay above the sensitivity limit (SNR$>$10) in each of the bands were included. The 
statistics for the field and source stars are listed in Table~\ref{fgstats}. As expected, 
the percentage of sources with an IR-excess is higher for source stars compared to field 
stars. Comparing these JHK colour-colour diagrams with similar diagrams for the entire LMC 
using 2MASS data (Nikolaev \& Weinberg~\cite{nikolaev}) shows that these field stars occupy 
the same regions in the colour-colour diagrams with a similar scatter around the mean values, 
although the giant branch stretches much further along the reddening vector in the diagrams 
for the entire LMC data set. 

\begin{figure}
   \centering
   \includegraphics[width=8.8cm]{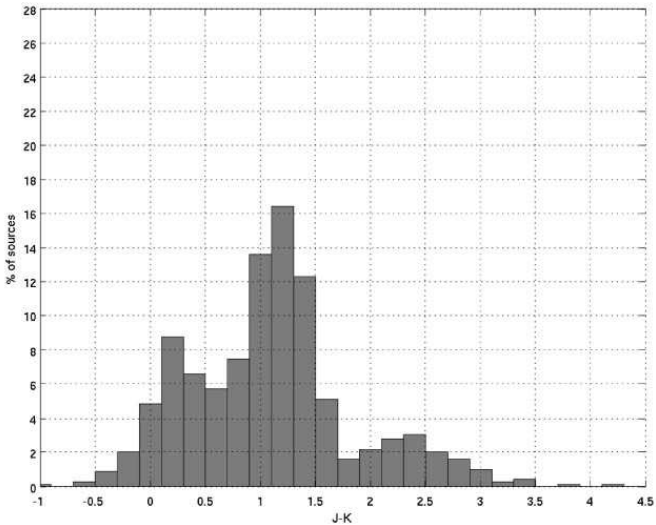}
   \caption{(J-K) colour distribution of JHK 2MASS data over the source. There is a peak
at (J-K)$\sim$0.35 which corresponds to the mean colour of stars earlier than G5 in the 2MASS
PSC (Finlator et al.~\protect\cite{finlator}). Another large peak occurs at (J-K)$\sim$1.
Redder sources with (J-K)$>$1.75 are also present. The median error in (J-K) is
$\sigma_{(J-K)}=0.08$. }
              \label{jhkChist}
    \end{figure}

\begin{figure}
   \centering
   \includegraphics[width=8.8cm]{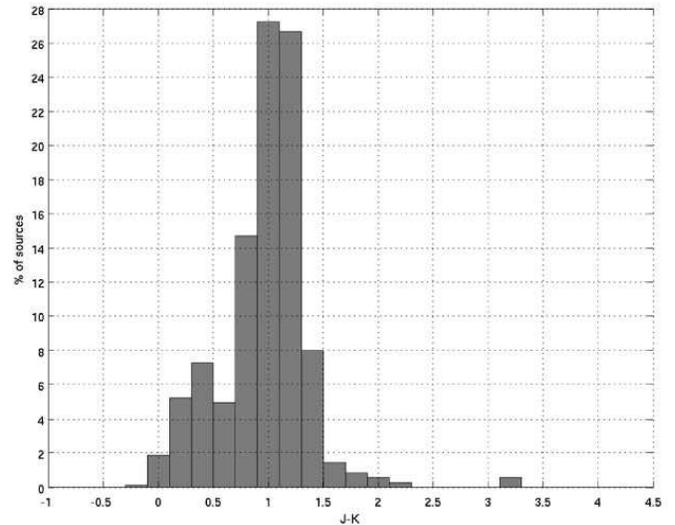}
   \caption{(J-K) colour distribution of the JHK 2MASS data for the field stars. The peak
at (J-K)$\sim$0.35 is still present, but is not as clear as in Fig.~\ref{jhkChist}.
The peak at (J-K)$\sim$1 is higher and red sources with (J-K)$>$1.4 are 
essentially missing. The median error in (J-K) is $\sigma_{(J-K)}=0.06$.}
              \label{jhkAhist}
    \end{figure}

\begin{figure}
   \centering
   \includegraphics[width=8.8cm]{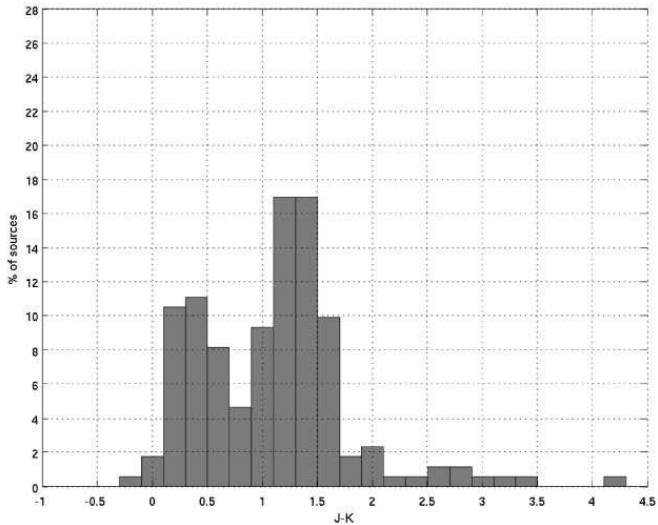}
   \caption{(J-K) distribution of the sources detected in all JHKL bands in 30 Doradus.
This distribution is similar to the distribution in Fig.~\ref{jhkChist}, with both peaks
at (J-K)$\sim$0.35 and (J-K)$\sim$1 and red sources, indicating that
the limits derived using the 2MASS JHK data also apply to the JHKL data. The median error
in (J-K) is $\sigma_{(J-K)}=0.05$.}
              \label{jhklhist}
    \end{figure}

As part of a comparison between the 2MASS data set and the Sloan
Digital Sky Survey, Finlator et al. (\cite{finlator}) trace stellar spectral sequences 
in JHK colour-colour diagrams derived from the 2MASS data. 
These show that stars of type M5 and later lie at (H-K$_S$)$\sim$0.2 and (J-H)$\sim$0.6. 
The sequence moves to bluer (H-K$_S$) and (J-H) colours for stars of type G5 and 
earlier, with (H-K$_S$)$\sim$0.05 and (J-H)$\sim$0.3 giving a (J-K) colour of $\sim$0.35. 
The distribution of (J-K) colours over the source (Fig.~\ref{jhkChist}) using the 2MASS 
data indeed shows a peak at (J-K)$\sim$0.35 with an additional peak at (J-K)$\sim$1 and a low rise 
in sources with even redder (J-K) colours at (J-K)$\sim2.4$. Both peaks can be found again in the 
same diagram for the field stars (Fig.~\ref{jhkAhist}), although here the peak at (J-K)$\sim$1 is 
higher and the very red sources are missing. The (J-K) distribution for the sources with
complete JHKL data (Fig.~\ref{jhklhist}) follows the same distribution as in Fig.~\ref{jhkChist} 
for the 2MASS data, indicating that (J-K) colour limits determined using the 2MASS JHK data are 
also applicable to the JHKL data. Assuming that the bulk of the stars in the off-source fields 
are foreground (i.e. not part of the 30 Doradus complex), sources bluer than the (J-K)=0.35 
colour limit are therefore likely to be foreground. 27 such sources are found, as indicated 
in Table~\ref{photres}. 12 of these have a moderate IR-excess ((K-L)$\lesssim0.6$) and 
three a slightly larger value (of (K-L)=1.1, 1.3 and 2.5 respectively). Excluding these stars
decreases the cluster disk fraction from 43\% to 42\% 
(see Sect.~\ref{IRind}) (Table~\ref{IRexcess}). Likely foreground stars are marked with boxes in 
Figs.~\ref{jhkl} to~\ref{CMD}.

\begin{table}
\caption{Statistics for the data from the 2MASS PSC for the combined off-source 
fields and the source field. The data for the field stars consists of 8 off-source
fields with a $3\arcmin$ radius each. The source data is a field centred on 30 Doradus
with a radius of $7\arcmin$. Only stars brighter than the sensitivity limit (SNR$>$10)
in all 3 bands (15.8, 15.1 and 14.3 for J, H and K respectively) were included.}
\label{fgstats}
\centering
\begin{tabular}{c c c c}
\hline\hline
Data & number of stars & number of IR-excess & fraction\\
& per arcmin$^2$ & sources per arcmin$^2$ & (in \%)\\
\hline
field & 3.03 & 0.17 & 6\\
source & 4.43 & 1.16 & 26\\
\hline
\end{tabular}
\end{table}

\section{Analysis}
\label{analysis}
\begin{table}
\caption{Number of detections in the different bands. The first column gives the total
number of detections in the SPIREX image. The second column gives the number of stars
that could be matched with the 2MASS PSC. Column three lists the number of stars 
additionally matched by comparison of the K- and L-band images. The last column lists the 
number of stars only found in the SPIREX L-band image. The second row lists the respective numbers
for stars brighter than the 90\% completeness limit. Using the (J-K) colour limit determined
in sec.~\ref{foreground} suggests 27 of the stars detected at JHKL are likely foreground stars.}
\label{phot}
\centering
\begin{tabular}{c c c c c}
\hline\hline
& Total & JHKL & KL & L\\
\hline
all stars & 215 & 184 & 7 & 24\\
$m_L<$13.5 (90\% limit) & 199 & 171 & 4 & 24\\
\hline
\end{tabular}
\end{table}

\subsection{Colour-colour and colour-magnitude diagrams}
Colour-colour and colour-magnitude diagrams were created using the 2MASS JHK-band
and the SPIREX L-band magnitudes. Only sources that were brighter than the 90\% completeness
limit in L-band were used in creating the diagrams and determining the fraction of IR-excess
sources. Using the intrinsic (V-K), (J-K), (H-K) and (K-L) colours of the main sequence and 
giant stars (Koornneef \cite{koorn}) and their absolute visual magnitudes M$_V$ (Allen \cite{allen}), 
the locations of the main sequence (spectral types O6-8 to M5) and the giant branch (spectral 
types K0 to M5) were plotted in all diagrams using a distance modulus of 18.7 magnitudes.

\begin{figure*}
   \centering
   \includegraphics[width=12cm]{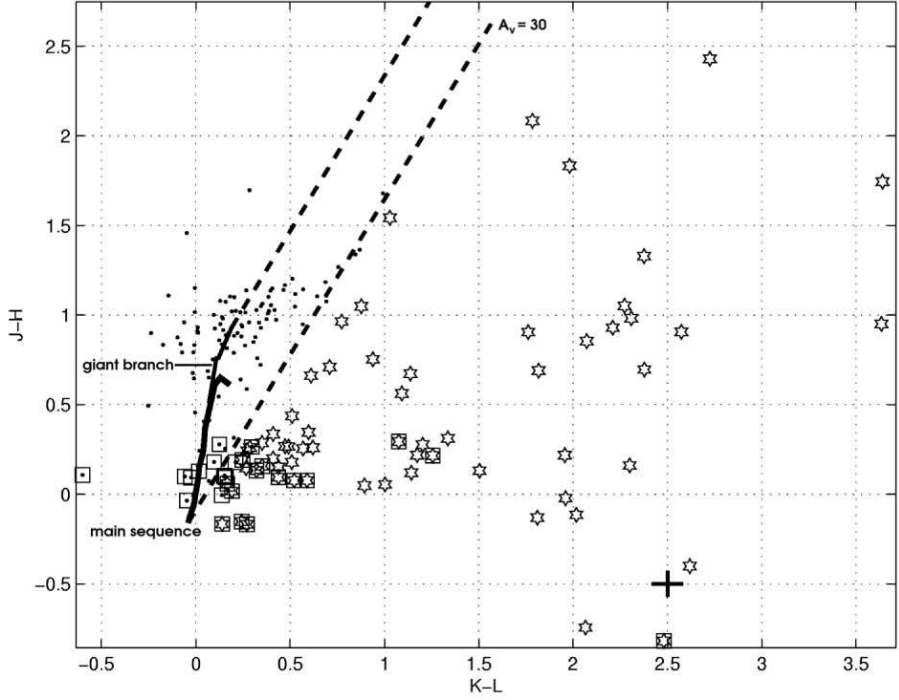}
   \caption{JHKL colour-colour diagram for 30 Doradus. The thick solid line is the 
main sequence for spectral types O6-8 to M5. The thin solid line is the giant 
branch for spectral types K0 ((J-H)=0.5, (K-L)=0.07) to M5 ((J-H)= 0.9, (K-L)=0.19), 
but is hard to see due to the number of stars in that region. Dashed lines are the 
reddening vectors up to \(A_{V}=30\). Star shaped symbols are sources identified as 
having an IR-excess. Sources in squares are likely to be foreground stars 
(Sec.~\ref{foreground}). Mean errors are indicated by the cross in the lower right of 
the diagram. 64 of the 171 stars detected in the JHKL bands lie outside the reddening 
band and are therefore counted as having an IR-excess. These are interpreted as coming from 
circumstellar disks around the stars (Sec.~\ref{IRind}).}
              \label{jhkl}
    \end{figure*}

\begin{figure*}
   \centering
   \includegraphics[width=12cm]{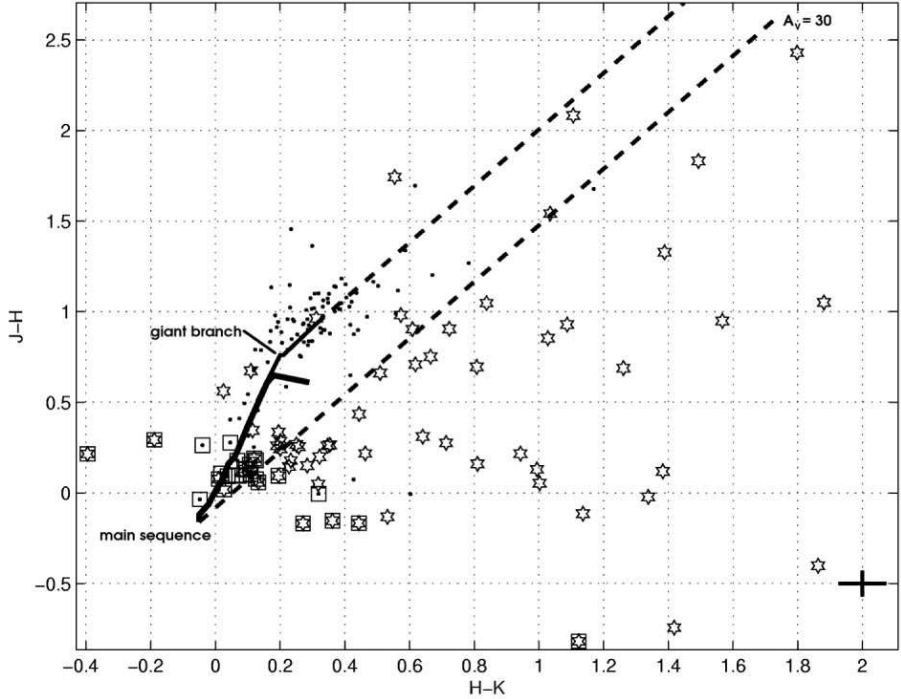}
   \caption{JHK colour-colour diagram for 30 Doradus. The same symbols are used as in 
Figure~\ref{jhkl}. Again the giant branch is hard to see, but extends from (H-K)=0.13 to (H-K)=0.31
with the same (J-H) values as in Figure~\ref{jhkl}. The diagram shows all sources detected in the 
JHK- and L-bands. The separation of stars with an IR-excess is much less clear in the JHK diagram than
in the JHKL diagram. Based on the JHK data for these stars alone would only give 49 
sources with IR-excess compared to 64 sources using the JHKL data (Table~\ref{IRexcess}) 
leading to an understimate of the cluster disk fraction.}
              \label{jhk}
    \end{figure*}

Figures \ref{jhkl} to \ref{CMD} show these colour-colour and colour-magnitude diagrams 
for 30 Doradus. In Figs. \ref{jhkl} and \ref{jhk} the thick solid curve shows 
the main-sequence. In Fig. \ref{CMD} the main sequence for stars ealier than type B0 
is represented by the thick solid curve. In all diagrams the thin solid curve shows the 
giant branch for types between K0 and M5. The dashed curve shows the reddening vector up 
to $A_{V}=30$ mags, assuming an extinction law $\propto {\lambda}^{-1.7}$.

\subsection{Fraction of reddened sources}
\begin{figure*}
   \centering
   \includegraphics[width=12cm]{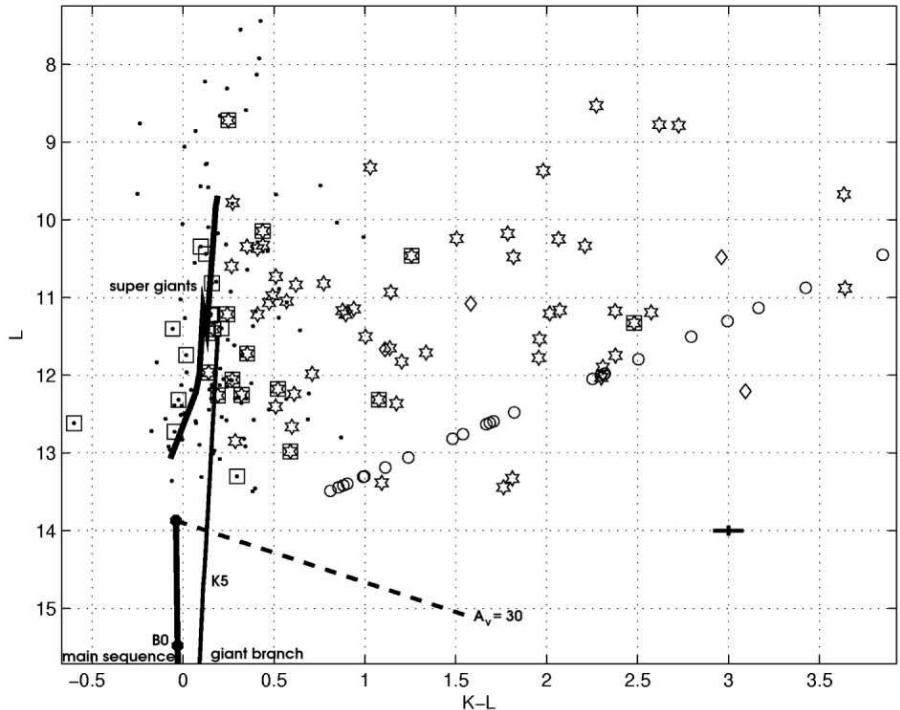}
   \caption{Infrared L vs (K-L) colour-magnitude diagram for 30 
Doradus. The lower thick solid line is the unreddened main sequence, the upper thick 
solid line shows the location of supergiants for a distance modulus of 18.7 magnitudes. 
The dashed line shows reddening up to \(A_{V}=30\) mags. Star shaped symbols are sources with 
an IR-excess as determined from the JHKL diagram. Diamond symbols are stars only found in 
K- and L- bands. Circle symbols are stars only found in the L-band and are located at the 
lower limit for their (K-L) colour. Sources in squares are likely to be foreground stars 
(Sec.~\ref{foreground}). The position of the stars only found in the KL-bands and L-band 
can be compared to the location of stars with an IR-excess (star shaped symbols). Those 
occupying the same region are also counted as having an IR-excess, giving an additional 4 
sources with an IR-excess found in K and L-bands and an additional 18 sources with an IR-excess 
found only in the L-band (Table~\ref{IRexcess}).}
              \label{CMD}
    \end{figure*}

A large number of stars in the JHKL diagram lie well to the right of the reddening band, 
indicating an IR-excess. In determining the IR-excess the individual errors of all stars
were used (Table~\ref{photres}). All stars which lie at least $1\sigma$ of their individual 
photometric error to the right and below the reddening band in the JHKL plane are defined 
to have an IR-excess and are marked with the star symbol. To estimate the uncertainty in 
this number, the number of stars that lie a $2\sigma$ distance to the right and below the 
reddening vector was also calculated. This procedure excludes 6 stars from the JHKL data set, 
one star from sources found only in K- and L-band and also one star from those seen at L-band 
only. The variation of the number of stars to the right of the reddening band when assuming  
$1\sigma$ and $2\sigma$ distances is taken as an estimate of the uncertainty of the number of 
stars that have an IR-excess. Stars defined to have an IR-excess in the JHKL plane are marked 
with the same symbol in Figs.~\ref{jhk} and~\ref{CMD} as well. In the colour-magnitude diagram 
(Fig.~\ref{CMD}) we can also include stars seen just at K and L (diamond shaped symbols).
We also include those seen only in the L-band by providing a lower limit on the (K-L) colour
(circle symbols) since the 2MASS sensitivity limit at K is 14.3 magnitudes. These stars
therefore lie to the right of the circles in these diagrams. This makes it possible to
also estimate what stars seen in just K- and L-bands have an IR-excess, by
comparing their location in Fig.~\ref{CMD} to the stars already identified as having an
IR-excess based on JHKL colours. Four of the stars seen only at K and L lie in the same region 
and are considered to have an IR-excess, and so are counted towards the total disk fraction. 
18 of the stars seen just in L-band are also located in this region and are also counted 
towards the total disk fraction. Table~\ref{IRexcess} lists these statistics. Excluding stars bluer 
than (J-K)=0.35, as discussed in Sect.~\ref{foreground}, decreases the fraction of reddened stars in 
Table~\ref{IRexcess} by $\sim$1\%.

\subsection{Mass range}
\begin{figure}
   \centering
   \includegraphics[width=8.8cm]{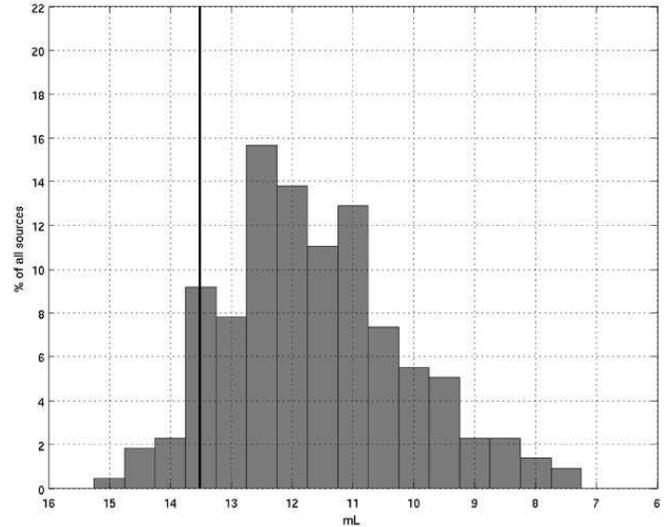}
   \caption{L band luminosity function of all stars detected at L-band in the SPIREX image of 
30 Doradus. Unreddened O3 main sequence stars have L-band magnitudes of 12.8 at the distance 
of 30 Doradus, unreddened O6-8 main sequence stars have magnitudes of $\sim13.9$ and 
unreddened B0 stars have L-band magnitudes of $\sim15.5$. The majority of detected stars have 
L-band magnitudes between 11 and 13, placing them where the massive, early type main sequence 
stars are. The vertical line at $m_L$=13.5 shows the 90\% completeness limit.}
              \label{alllum}
    \end{figure}

\begin{figure}
   \centering
   \includegraphics[width=8.8cm]{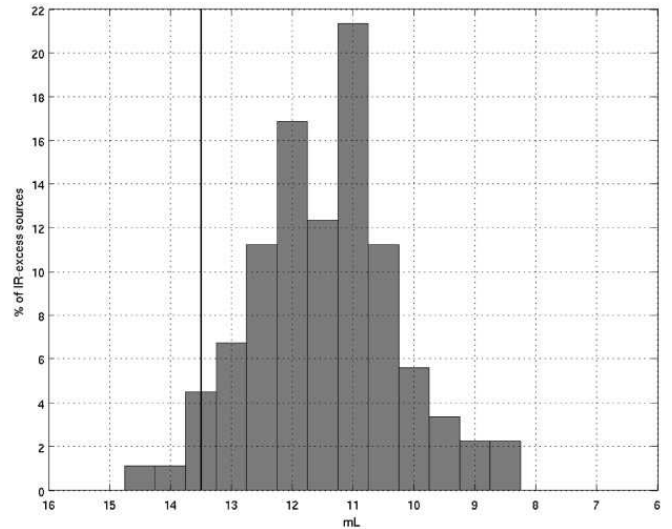}
   \caption{The percentage of sources with an IR-excess in the different magnitude 
intervals. The distribution peaks at an L-band magnitude of $\sim11$ indicating high mass stars. 
IR-excess is however detected down to magnitudes of L$\sim14.5$. The vertical line at $m_L$=13.5 
shows the 90\% completeness limit.}

              \label{irlum}
    \end{figure}

Figure~\ref{alllum} shows the number of detected sources vs L-band magnitude for all sources 
in the SPIREX image. At the distance of 30 Doradus O3 main sequence stars have L-band 
magnitudes of 12.8, O6-8  main sequence stars have magnitudes of $\sim13.9$ and B0 main
sequence stars have L-band magnitudes of $\sim15.5$. Masses can be crudely estimated by 
comparing the magnitudes with those for stars of known spectral type (Allen~\cite{allen}).
The histogram in Fig.~\ref{alllum} peaks at m$_L \sim12$ magnitudes, showing that we are 
mainly picking up the massive, early main sequence stars, extending down to masses of order 
20 M$_{\sun}$ (ie late-type O stars). Figure~\ref{irlum} shows the
percentage of stars in each magnitude interval that have an IR-excess. The distribution seems 
to peak closer to m$_L \sim11$ magnitudes, suggesting that the IR-excess possibly is not independent 
of mass. However, the IR-excess biases these estimates to higher mass. IR-excess is 
detected for masses down to $\sim$23 M$_{\sun}$, confirming that 30 Doradus is a region of 
high mass star formation.

\begin{table*}
\caption{Number of stars found with IR-excess and the reddening fraction. Numbers in
the JHKL, KL and L columns are the number of stars with IR-excess found in the
respective bands. These are listed  assuming 1$\sigma$ and 2$\sigma$ distances from the 
reddening band. Columns 7 and 8 give the total number of sources with IR-excess at 
1$\sigma$ and 2$\sigma$. Column 9 gives the cluster disk fraction. Excluding possible
foreground stars decreases the CDF to $\sim$42\%. Only sources that are brighter than the
90\% completeness limit (13.5 mag in L-band) are included when calculating the fraction of
IR-excess sources.}
\label{IRexcess}
\centering
\begin{tabular}{c c c c c c c c c}
\hline\hline
JHKL (1$\sigma$) & JHKL (2$\sigma$) & KL (1$\sigma$) & KL (2$\sigma$) & L (1$\sigma$) & L (2$\sigma$) & total (1$\sigma$) & total (2$\sigma$) & frac\\
\hline
64 & 58 & 4 & 3 & 18 & 17 & 86 & 78 & 43$\pm$5\%\\
\hline
\end{tabular}
\end{table*}

\section{Discussion}
\label{discussion}
\subsection{IR-excess as an indicator of circumstellar disks}
\label{IRind}
Using the JHKL data shows that $\sim43\pm5\%$ of the stars detected in 30 Doradus
at L-band lie to the right of the reddening band defined by interstellar 
extinction, and thus have an IR-excess over a stellar photosphere. Models for sources with 
circumstellar disks can explain such positions in the colour-colour diagrams (eg. Lada 
\& Adams~\cite{ladaadams}). Stars showing an IR-excess in the colour-colour and 
colour-magnitude diagrams (Table~\ref{IRexcess}) are interpreted as having circumstellar 
disks. The cluster disk fraction (CDF) is therefore estimated as the fraction of stars 
with IR-excess.

\subsection{Spatial Distribution of IR-excess sources}
\label{distribution}
\begin{figure*}
   \centering
   \includegraphics[width=12cm]{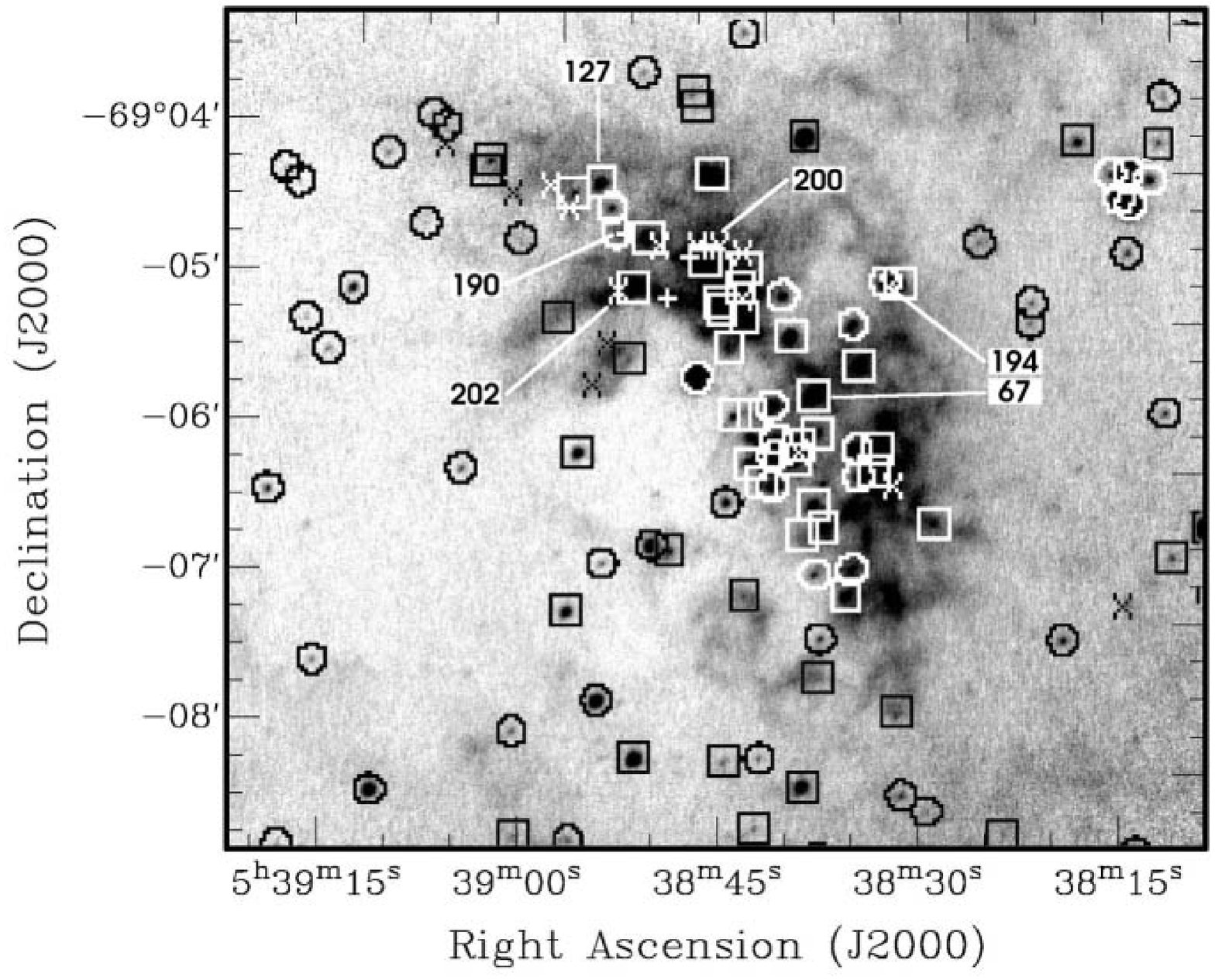}
   \caption{Spatial distribution of sources in 30 Doradus. Circles mark stars
without an IR-excess, squares mark stars with an IR-excess found in the 
JHKL-bands, plus signs mark stars with an IR-excess found only in the KL-bands
and crosses mark stars with an IR-excess only found in L-band (there is no difference between 
white/black sources, this is only in order to make the stars in the nebulous regions more visible). 
A string of stars only detected in the L-band follows the nebulosity to the north-east of the 
central region. Stars with an IR-excess are confined to the inner regions inside the nebula. Sources
without an IR-excess are mainly distributed in regions without any nebulosity. The 6 reddest
stars ((K-L)$>3$) are marked in the image with their id from Table~\ref{photres}.}
              \label{spafig}
    \end{figure*}

Figure~\ref{spafig} shows the spatial distribution of the detected stars in the
central part of the SPIREX image (see also Fig.~\ref{colour}). It is apparent that 
stars that do not show an IR-excess lie outside the nebulous regions, whereas 
IR-excess stars lie within the central nebula and along the arcs and knots in the 
image. Sources not detected in the J- and H-bands are essentially confined to the 
string of nebulous regions in the centre. There seems to be a string of stars only 
detected in the L-band along the arc north of the central cavity. The centre of 30
Doradus itself lies in a halo of molecular gas (Garay et al.~\cite{garayetal}) and the
sources only found in L-band are found to preferentially lie in the regions where 
[CII] and CO line emission is detected (Poglitsch et al.~\cite{poglitschetal}).
IR-excess sources detected in all four bands tend to spread out further throughout the 
region.

\subsection{Disk evolution}
\label{diskevo}
Together with the CDF and the age of the cluster, it is possible to estimate 
the lifetime of circumstellar disks. Walborn and Blades (\cite{walbornblades}) find
several regions with different ages in the 30 Doradus region. The massive 
central region is estimated to be $\sim2-3$ Myr old, but a much younger 
population of less than 1 Myr and an older population of about 4-6 Myr are also 
found (Walborn \& Blades \cite{walbornblades}; Massey \& Hunter 
\cite{masseyhunter}). Since young stars are very bright in the IR, sources with an
IR-excess can be intepreted as being young stellar objects with some of these sources 
possibly being more deeply embedded protostars surrounded by dust shells rather 
than disks (Sect~\ref{highmass}). However, depending on the distribution of dust around 
evolved stars, these also present an explanation for the observed IR-excess and it would 
be necessary to take spectra of the sources to ascertain their evolutionary state. 

Adopting a mean age for 30 Doradus of 2-3 Myr, the number of stars with 
IR-excess ($\sim$42\%; Table~\ref{IRexcess} excluding foreground
stars) would imply that more than 50\% of the circumstellar disks have disappeared 
after 2-3 Myr. It is possible that the intense radiation from the early type stars 
in the 30 Doradus region effectively destroys circumstellar disks
and therefore decreases the CDF. Observations of HII regions in the Galaxy have 
shown that externally illuminated circumstellar disks get photoevaporated and 
disappear after 0.01-0.1 Myr. This has for example been seen in the proplyds in the
Orion Nebula (O'Dell \& Wen \cite{odellwen}). Proplyds are disks or flattened 
envelopes of circumstellar material that are photoevaporated from the outside by an
external ionizing field. The variation in the form of proplyds is explained
by a balance between stellar gas pressure and radial pressure and radiation pressure
from dominant stars in the region. If this is the case for 30 Doradus, the lifetime for
disks estimated using the CDF of this region would be too low, and derived 
lifetimes a lower limit. 

\begin{figure*}
   \centering
   \includegraphics[width=12cm]{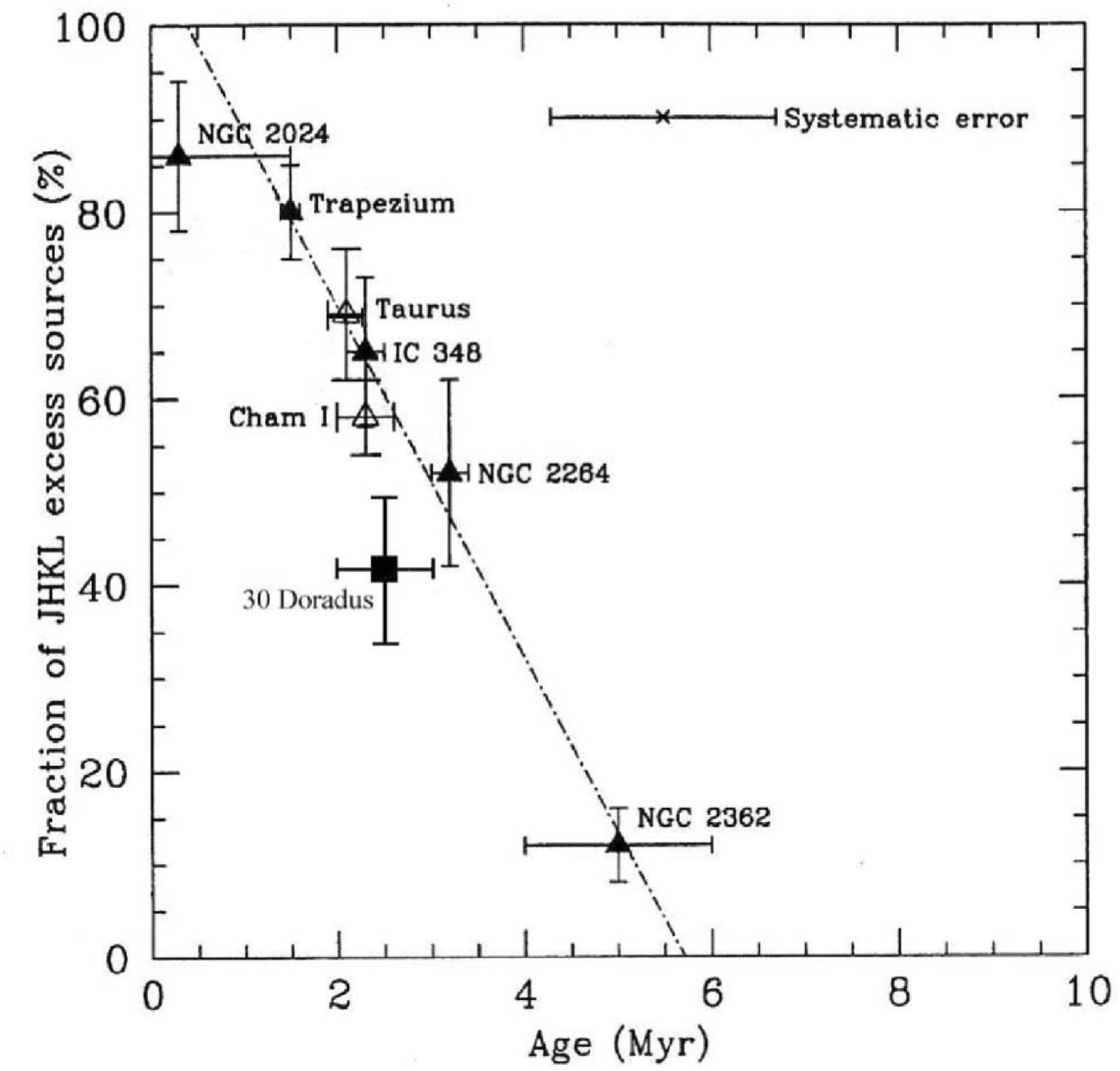}
   \caption{Cluster disk fraction (CDF) for 6 clusters by Haisch et al. \cite{haischetalb}
 vs their mean ages. The CDFs were determined using JHKL data as described in their paper.
The location of 30 Doradus is indicated by a square with error bars indicating the 
uncertainty in age and disk fraction. The dot-dashed line is the best fit determined by 
Haisch et al. (\protect\cite{haischetalb}).}
              \label{CDFfig}
    \end{figure*}

Combined with earlier JHKL observations (Haisch et al. \cite{haischetalb}) of 
clusters NGC 2264, NGC 2362, NGC 1960 (ages between 2.5-30 Myr) and younger clusters
NGC 2024, Trapezium and IC 348, with ages down to 0.3 Myr (NGC 2024), predictions
on the lifetime of circumstellar disks can be made. For all these clusters the 
fraction of sources with JHKL IR-excess is determined. Each cluster is then plotted in 
a CDF vs. age diagram (Fig.~\ref{CDFfig}). The position of 30 Doradus in the
diagram is indicated. The error in the CDF for 30 Doradus is $\pm8\%$ due to the 
uncertainty in the number of IR-excess sources ($\pm5\%$, Table~\ref{IRexcess}) and 
allowing for an uncertainty in the number of foreground stars. The errors in the age 
reflect a mean age of 2-3 Myr. The location of 30 Doradus is consistent with the fit of 
Haisch et al., despite this being for sources which extend to much lower masses 
(0.13M$_{\sun}$ for NGC 2024 up to 1M$_{\sun}$ for NGC2362 at the completeness limit; 
Haisch et al.~\cite{haischetalb}) than our study of 30 Doradus.

A least-squares straight line fit to the data shows a linear relationship 
between disk fraction and cluster age. An estimated disk lifetime of 6 Myr is 
derived. The initial disk fraction appears to be very high ($\sog$ 80\%) and then 
declines linearly with increasing age. Taking the contamination of foreground
stars into consideration yields a CDF of $\sim$42\% and an age of $\sim2.5$Myr, 
then the CDF derived here lies just below that which Haisch et al. would predict
(Fig.~\ref{CDFfig}). This indicates a quicker evolution of circumstellar disks around 
massive stars or that the disks indeed are being destroyed due to photoevaporation or 
possibly both.

\subsection{High mass stars}
\label{highmass}
Whereas 30 Doradus contains a large range of stellar masses (Brandner 
\cite{brandner}), our observations are only sensitive to the most massive stars.
Since the derived CDF lies at the lower end of the Haisch et al. prediction, this 
again indicates a faster evolution of circumstellar disks around massive stars.
Previous studies of the 30 Doradus complex show 20 sources with IR-excess in JHK 
colour-colour diagrams (Brandner \cite{brandner}). 8 of these could be matched with
L-band sources found in the SPIREX image (Table~\ref{photres}) and have an IR-excess. The 
20 sources were associated with Class I protostars and Herbig Ae/Be or T Tauri stars by 
comparing the measured colours with models. This again confirms the connection between 
star formation, circumstellar disks and IR-excess. The majority of sources in the Brandner
paper however lie below the 2MASS detection limit (15.8 mags in J-band), which 
explains why we only identify 8 sources from their work. A scatter in the photometry of
individual sources can be explained by differences in spatial resolution and nonstandard
NICMOS passbands. In the Brandner study the sources of star formation were found to follow
the densest regions of the molecular gas and the interface with the hollowed-out region 
from the central stars. The same is true for most of the 24 stars detected only in the 
L-band, which lie in the nebulous region north of the central star cluster 
(Fig.~\ref{spafig}) (see Sect.~\ref{distribution}). In the L vs (K-L) colour-magnitude 
diagram they occupy the same region as the IR-excess stars from the (J-H) vs (K-L) 
colour-colour diagram. This makes it likely that these stars are in the process of forming
and also contain circumstellar disks. These properties are characteristic of protostellar 
objects and the sources only detected in the L-band are possibly the equivalent of 
massive Class I protostars (Lada et al. \cite{ladaetal}). Since these stars are only 
detected in the L-band, and not at shorter wavelengths, it also suggests they are heavily 
embedded in dust shells, and are massive young stellar objects.

\subsection{The reddest sources}
In Fig.~\ref{spafig} six particularly red sources ((K-L)$>3$) are marked with their id from 
Table~\ref{photres}. Two of these are found in all four bands, one only in K and L and three
are only found in L-band. The reddest source (\#194, (K-L)$>$3.8) lies in a knot to the north west
of the central cluster. The other sources are in the central nebula and coincide with the brightest 
arcs and knots of the nebula. Three of the sources follow the chain of stars only detected in L-band 
towards the north-east of the image (\#127, \#190 and \#200). Source \#67 lies close to the 
central cluster.

\subsection{Supergiants}
The majority of the stars in the colour-magnitude diagram for 30 Doradus 
(Fig.~\ref{CMD}) are located above the main sequence, meaning that these stars 
are brighter than O6-8 type stars (the sensitivity threshold of the SPIREX images
lies at 13.5 magnitudes, $\sim$0.4 mag  brighter than the location of O6-8 stars). 
Massey and Hunter (1998) detected 31 O8V or earlier type stars in the
30 Doradus cluster. Other studies also find a significant number of supergiants and 
giants (Walborn \& Blades \cite{walbornblades}; Bosch et al. \cite{boschetal}; 
Parker \cite{parker}). The location of supergiants between B0 and M0 is drawn in 
the 30 Doradus CMD in Fig.~\ref{CMD} as the upper thick solid line above the 
main sequence. Of the four papers on the stellar population quoted here, 
unfortunately, only Parker gives coordinates, so that it is not possible to match
our data with the other three studies. 

Matching a B0I supergiant and a O3V star from the Parker paper with the SPIREX data
(stars \#76 and \#187, see Table~\ref{photres}) shows that these two stars indeed occupy 
the region of the CMD where most of the detected stars lie, making it likely that
supergiant, giant and early type OV stars are present in 30 Doradus.

\section{Conclusions}
L-band photometry for 30 Doradus has been presented. Combining the L-band data with 
shorter wavelength data from the JHK bands proves to be a powerful method to detect
IR-excess. In particular, this IR-excess can be interpreted as emission from
circumstellar disks, making L-band observations an essential tool for identifying 
these disks. The clearer separation of stars in the JHKL diagrams provides a better 
estimate of the total cluster disk fraction than JHK data alone does. Only using 
JHK data tends to underestimate the number of stars with IR-excess (an analysis using
the same stars as in the JHKL colour-colour diagram, but only considering the JHK
information, gives only 49 IR-excess sources as compared to 64 when using the JHKL data
(Table~\ref{IRexcess})). Comparing the results obtained for 30 Doradus with other L-band 
studies of star forming regions in the Galaxy shows an inverse relation between age and 
disk fraction, giving further support to an estimated disk lifetime of ~6 Myr and a very 
high initial disk fraction. Since the sensitivity limit achieved for the observations in 
this study only includes stars of type O6 and brighter, these results indicate that 
high-mass stars form in a similar way to low-mass stars through circumstellar disks,
however, it is possible that disks around high mass stars evolve faster than around low
mass stars. Several sources are seen in the L-band image that are not found at shorter 
wavelengths. These sources are possibly heavily embedded, high-mass protostars. The 
positions of these stars in the colour-magnitude diagram indicate that they might 
also be circumstellar disk systems. It is also possible that the IR-excess originates
from evolved stars (AGB stars) or young high and intermediate mass stars 
(see Sect.~\ref{diskevo}). In particular in the case of a complex environment with 
mutiple generations as 30 Doradus (see Sect.~\ref{30dor}), a closer spectral examination
is necessary to ascertain the evolutionary state of the sources.

No other extensive investigation of the JHKL colour excess has been made for 
30 Doradus, making this the first study of infrared excess using L-band data for a 
significant number of stars for this region. The data are consistent with similar data 
obtained from Galactic sources, indicating that similar processes are at work in star 
formation in the LMC.

The data illustrates the ability of Antarctic telescopes, where the low 
thermal background improves IR sensitivity, to probe the state of the 
embedded stars in star forming regions.

\begin{acknowledgements}
This work could not have been conducted without the great help
received from many colleagues within the US CARA and the Australian
JACARA organisations whose efforts made the SPIREX/Abu project at the
South Pole such a success, to whom we are extremely grateful.  We also
acknowledge the funding support from the Australian Research Council
and the Australian Major National Research Facilities program that
made this work possible in the first place. 
We thank the referee, Wolfgang Brandner, for his perceptive comments, which have 
greatly improved this paper. 
This publication makes use of data products from the Two Micron 
All Sky Survey, which is a joint project of the University of Massachusetts and the Infrared 
Processing and Analysis Center/California Institute of Technology, funded by the National 
Aeronautics and Space Administration and the National Science Foundation. 
This research has made use of the NASA/ IPAC Infrared Science Archive, which is operated by 
the Jet Propulsion Laboratory, California Institute of Technology, under contract with the 
National Aeronautics and Space Administration.
\end{acknowledgements}

\begin{longtable}{c c c c c c c c c c c c}
\caption{\label{photres}(Also available in electronic form at the CDS). Magnitudes for all 
sources (including foreground sources and sources below the 90\% completeness limit) in 30 Doradus.
 Stars with measurements in all four bands are listed first. Then stars with measurements in K and 
L only, and finally stars just detected in L-band. Column 1 gives the source id. Columns. 2 and 3 
the RA and Dec respectively in J2000. 
The coordinates for sources found in all four bands are from the point source catalogue, 
positions for the remaining stars are determined from reference 2MASS images. 
Columns 4, 6, 8 and 10 give the JHK- and L-band magnitudes respectively. Columns 5, 7, 9 and
11 give the photometric errors. For sources detected in all bands, the JHK magnitudes and errors
are taken from the 2MASS PSC. A `null' as error indicates that no photometric error was given
in the PSC. The L-band errors are combined from the errors in daophot and the errors due to the
zero point correction. Sources only detected in the K- and L-bands have magnitudes from this work. 
For sources not detected at J,H or K the upper limits on these magnitudes are 15.8, 15.1 and 14.3 
respectively. Stars with an IR-excess are marked with an `e' in Col. 12 (comments). 
Likely foreground stars are marked with `fg'. Stars that matched sources in Rubio et al. 
(\cite{rubio}), Brandner et al. (\cite{brandner}) and Parker at al (\cite{parker}) are 
marked with `Rubio', `Brandner' and `Parker' respectively together with the id assigned 
in the respective papers.}\\
\hline\hline
id & RA (J2000) & Dec (J2000) & $m_{J}$ & $\sigma_{J}$ & $m_{H}$ & $\sigma_{H}$ & $m_{K}$ & $\sigma_{K}$ & $m_{L}$ & $\sigma_{L}$ & comments\\
& (h m s) & (d m s)& & & & & & & & &\\
\hline
\endfirsthead
\caption{continued.}\\
\hline
id & RA (J2000) & Dec (J2000) & $m_{J}$ & $\sigma_{J}$ & $m_{H}$ & $\sigma_{H}$ & $m_{K}$ & $\sigma_{K}$ & $m_{L}$ & $\sigma_{L}$ & comments\\
& (h m s) & (d m s)& & & & & & & & &\\
\hline
\endhead
\hline
\endfoot
1   & 5 37 50.2 & -69 04 24.2 & 10.0 & 0.02 & 09.5 & 0.02 & 09.4 & 0.02 & 09.7 & 0.05 & \\
2   & 5 37 50.5 & -69 04 02.3 & 14.7 & 0.04 & 12.8 & 0.03 & 11.4 & 0.02 & 09.4 & 0.05 & e \\
3   & 5 37 54.7 & -69 09 03.2 & 10.3 & 0.02 & 09.4 & 0.02 & 09.1 & 0.02 & 09.1 & 0.08 & \\
4   & 5 37 57.0 & -69 03 38.9 & 12.9 & 0.02 & 11.9 & 0.03 & 11.6 & 0.03 & 10.8 & 0.05 & e \\
5   & 5 37 58.1 & -69 05 03.8 & 12.6 & 0.02 & 11.6 & 0.03 & 11.2 & 0.02 & 10.9 & 0.05 & \\
6   & 5 37 59.1 & -69 08 41.3 & 11.0 & 0.02 & 10.0 & 0.03 & 09.7 & 0.02 & 09.6 & 0.05 & \\
7   & 5 38 05.5 & -69 09 26.6 & 13.2 & 0.02 & 12.2 & 0.03 & 11.8 & 0.02 & 11.3 & 0.12 & \\
8   & 5 38 05.6 & -69 09 09.7 & 15.1 & 0.06 & 14.3 & 0.06 & 13.2 & 0.04 & 11.2 & 0.10 & e \\
9   & 5 38 06.2 & -69 05 42.0 & 14.1 & 0.02 & 13.1 & 0.03 & 12.9 & 0.04 & 12.3 & 0.06 & \\
10  & 5 38 06.2 & -69 02 35.5 & 11.8 & 0.03 & 10.8 & 0.03 & 10.2 & 0.02 & 10.1 & 0.04 & \\
11  & 5 38 06.6 & -69 03 45.4 & 10.5 & 0.02 & 09.7 & 0.02 & 09.4 & 0.02 & 09.3 & 0.04 & \\
12  & 5 38 08.8 & -69 03 45.4 & 13.5 & 0.02 & 12.6 & 0.03 & 12.4 & 0.03 & 12.1 & 0.07 & \\
13  & 5 38 08.9 & -69 06 47.9 & 14.3 & 0.03 & 13.2 & 0.04 & 12.9 & 0.03 & 12.2 & 0.06 & \\
14  & 5 38 09.6 & -69 06 21.2 & 09.3 & 0.02 & 09.1 & 0.02 & 09.0 & 0.02 & 08.7 & 0.04 & e, fg \\
15  & 5 38 10.0 & -69 03 43.2 & 12.1 & 0.02 & 10.9 & 0.03 & 10.2 & 0.02 & 09.7 & 0.04 & \\
16  & 5 38 12.0 & -69 06 34.2 & 12.8 & 0.02 & 12.7 & 0.03 & 12.6 & 0.03 & 12.3 & 0.06 & e, fg \\
17  & 5 38 12.7 & -69 02 54.2 & 14.4 & 0.03 & 13.4 & 0.03 & 13.2 & 0.03 & 12.8 & 0.08 & \\
18  & 5 38 13.2 & -69 05 36.6 & 12.7 & 0.02 & 12.3 & 0.03 & 12.2 & 0.03 & 12.2 & 0.05 & \\
19  & 5 38 13.6 & -69 08 33.0 & 15.3 & 0.05 & 14.0 & 0.04 & 13.7 & 0.05 & 12.8 & 0.10 & \\
20  & 5 38 14.6 & -69 00 57.6 & 12.0 & 0.02 & 10.9 & 0.02 & 10.6 & 0.02 & 10.3 & 0.05 & \\
21  & 5 38 14.8 & -69 03 29.9 & 14.0 & 0.04 & 13.3 & 0.04 & 13.0 & 0.04 & 12.9 & 0.09 & \\
22  & 5 38 14.9 & -69 03 48.6 & 12.5 & 0.02 & 12.5 & 0.03 & 12.5 & 0.02 & 12.3 & 0.06 & e, fg \\
23  & 5 38 15.4 & -69 04 03.4 & 11.9 & 0.04 & 11.7 & 0.04 & 11.6 & 0.04 & 11.4 & 0.05 & fg \\
24  & 5 38 15.7 & -69 04 37.2 & 16.0 & 0.10 & 15.6 & 0.18 & 15.3 & 0.22 & 14.5 & 0.30 & \\
25  & 5 38 16.0 & -69 10 11.3 & 09.3 & 0.02 & 08.3 & 0.01 & 07.9 & 0.02 & 07.4 & 0.04 & \\
26  & 5 38 16.7 & -69 04 14.2 & 09.8 & 0.04 & 08.9 & 0.03 & 08.6 & 0.03 & 08.3 & 0.04 & \\
27  & 5 38 16.7 & -69 04 33.2 & 13.5 & 0.02 & 12.6 & 0.02 & 12.3 & 0.02 & 12.1 & 0.05 & \\
28  & 5 38 17.0 & -69 04 00.8 & 10.1 & 0.03 & 09.2 & 0.03 & 08.9 & 0.02 & 08.7 & 0.04 & \\
29  & 5 38 17.6 & -69 04 12.0 & 10.8 & 0.03 & 10.0 & 0.03 & 09.7 & 0.02 & 09.6 & 0.04 & \\
30  & 5 38 18.3 & -69 04 02.3 & 12.4 & 0.05 & 12.3 & 0.05 & 12.3 & 0.06 & 12.3 & 0.07 & fg \\
31  & 5 38 19.9 & -69 07 10.2 & 13.1 & 0.02 & 12.3 & 0.02 & 12.1 & 0.03 & 11.7 & 0.05 & \\
32  & 5 38 20.9 & -69 03 50.0 & 12.2 & 0.02 & 11.8 & 0.03 & 11.6 & 0.03 & 11.2 & 0.05 & e \\
33  & 5 38 21.4 & -69 09 49.0 & 13.2 & 0.02 & 12.1 & 0.02 & 11.8 & 0.02 & 11.4 & 0.07 & \\
34  & 5 38 23.6 & -69 08 29.4 & 15.1 & 0.05 & 14.5 & 0.05 & 14.5 & 0.11 & 13.4 & 0.12 & e \\
35  & 5 38 23.7 & -69 04 55.6 & 13.8 & 0.03 & 12.8 & 0.03 & 12.5 & 0.04 & 12.5 & 0.06 & \\
36  & 5 38 23.7 & -69 05 03.5 & 12.6 & 0.02 & 12.6 & 0.02 & 12.7 & 0.02 & 12.7 & 0.06 & fg \\
37  & 5 38 26.7 & -69 08 52.8 & 09.8 & 0.02 & 08.8 & 0.03 & 08.3 & 0.02 & 07.9 & 0.04 & \\
38  & 5 38 27.8 & -69 06 14.8 & 15.2 & 0.05 & 14.3 & 0.05 & 14.2 & 0.07 & 14.3 & 0.25 & \\
39  & 5 38 27.9 & -69 04 32.2 & 14.8 & 0.08 & 13.6 & 0.06 & 13.3 & 0.05 & 12.6 & 0.06 & \\
40  & 5 38 29.5 & -69 08 21.5 & 15.2 & 0.06 & 14.0 & 0.05 & 13.9 & 0.05 & 13.5 & 0.15 & \\
41  & 5 38 29.9 & -69 11 16.4 & 12.4 & 0.02 & 11.2 & 0.03 & 10.9 & 0.02 & 10.4 & 0.07 & \\
42  & 5 38 30.0 & -69 06 25.9 & 13.6 & 0.02 & 12.9 & 0.02 & 12.8 & 0.02 & 11.7 & 0.10 & e \\
43  & 5 38 31.0 & -69 01 16.0 & 10.9 & 0.02 & 10.6 & 0.02 & 10.6 & 0.03 & 10.4 & 0.05 & fg \\
44  & 5 38 31.3 & -69 08 16.1 & 13.7 & 0.04 & 12.8 & 0.04 & 12.6 & 0.04 & 12.3 & 0.05 & \\
45  & 5 38 31.7 & -69 02 14.6 & 14.6 & 0.10 & 13.6 & 0.09 & 12.5 & 0.07 & 10.3 & 0.08 & e \\
46  & 5 38 32.0 & -69 07 44.0 & 14.9 & null & 14.7 & 0.10 & 13.7 & 0.07 & 11.8 & 0.06 & e \\
47  & 5 38 33.6 & -69 04 50.5 & 12.1 & 0.04 & 11.8 & 0.07 & 11.5 & 0.06 & 10.8 & 0.04 & e \\
48  & 5 38 34.6 & -69 05 56.8 & 13.7 & 0.14 & 12.7 & 0.10 & 10.8 & 0.04 & 08.5 & 0.04 & Brandner 3a, e \\
49  & 5 38 34.7 & -69 06 06.1 & 13.6 & 0.14 & 13.5 & 0.23 & 12.5 & 0.12 & 11.5 & 0.07 & e \\
50  & 5 38 34.8 & -69 04 50.2 & 13.3 & 0.07 & 12.1 & 0.05 & 11.6 & 0.05 & 11.1 & 0.05 & \\
51  & 5 38 36.0 & -69 06 09.0 & 12.1 & 0.05 & 12.0 & 0.07 & 12.0 & 0.06 & 12.6 & 0.11 & fg \\
52  & 5 38 36.1 & -69 06 46.8 & 12.4 & 0.05 & 12.2 & 0.05 & 12.0 & 0.07 & 12.0 & 0.06 & \\
53  & 5 38 36.1 & -69 05 57.8 & 10.7 & 0.02 & 10.6 & 0.02 & 10.4 & 0.02 & 10.3 & 0.04 & fg \\
54  & 5 38 36.4 & -69 06 57.2 & 12.1 & 0.03 & 11.8 & 0.03 & 11.5 & 0.03 & 11.0 & 0.04 & e \\
55  & 5 38 36.6 & -69 05 26.5 & 14.1 & null & 14.9 & 0.39 & 13.8 & 0.22 & 11.3 & 0.07 & e, fg \\
56  & 5 38 37.0 & -69 05 07.8 & 11.6 & 0.03 & 11.5 & 0.04 & 11.4 & 0.04 & 11.2 & 0.06 & fg \\
57  & 5 38 37.4 & -69 08 42.7 & 13.4 & 0.02 & 12.5 & 0.03 & 12.3 & 0.02 & 12.2 & 0.05 & \\
58  & 5 38 38.0 & -69 07 30.0 & 15.8 & null & 14.8 & null & 14.2 & 0.20 & 11.9 & 0.06 & e \\
59  & 5 38 38.1 & -69 07 14.9 & 14.9 & 0.04 & 13.2 & 0.03 & 12.6 & 0.03 & 12.3 & 0.06 & \\
60  & 5 38 38.5 & -69 06 29.9 & 13.9 & 0.10 & 12.9 & null & 12.0 & null & 11.2 & 0.09 & Rubio 98, e \\
61  & 5 38 38.7 & -69 06 13.0 & 14.2 & 0.08 & 14.0 & 0.11 & 13.7 & null & 14.1 & 0.16 & \\
62  & 5 38 38.8 & -69 06 49.3 & 12.7 & 0.02 & 12.4 & null & 12.3 & 0.02 & 12.1 & 0.06 & \\
63  & 5 38 38.9 & -69 08 14.6 & 11.2 & 0.03 & 10.9 & 0.03 & 10.7 & 0.03 & 10.3 & 0.04 & e \\
64  & 5 38 39.1 & -69 06 21.2 & 11.7 & null & 11.8 & 0.07 & 11.5 & 0.05 & 11.2 & 0.05 & e, fg \\
65  & 5 38 39.3 & -69 05 52.1 & 14.0 & 0.10 & 13.7 & 0.12 & 13.0 & 0.07 & 11.7 & 0.05 & Rubio 118, e \\
66  & 5 38 39.4 & -69 06 06.1 & 13.5 & 0.12 & 13.4 & 0.19 & 13.3 & 0.13 & 14.2 & 0.20 & \\
67  & 5 38 39.7 & -69 05 38.8 & 15.8 & 0.29 & 14.9 & 0.11 & 13.3 & 0.05 & 09.7 & 0.04 & e \\
68  & 5 38 39.8 & -69 06 34.9 & 14.8 & 0.11 & 14.8 & 0.21 & 13.5 & 0.11 & 11.5 & 0.06 & e \\
69  & 5 38 40.2 & -69 09 33.1 & 14.7 & 0.04 & 13.9 & 0.05 & 13.6 & 0.05 & 13.7 & 0.16 & \\
70  & 5 38 40.5 & -69 05 57.1 & 11.2 & 0.09 & 11.1 & 0.11 & 10.8 & 0.07 & 10.3 & 0.04 & e \\
71  & 5 38 40.8 & -69 06 02.9 & 11.5 & null & 11.3 & null & 11.7 & 0.14 & 10.5 & 0.04 & e, fg \\
72  & 5 38 40.9 & -69 10 08.8 & 13.5 & 0.02 & 12.7 & 0.02 & 12.5 & 0.02 & 12.5 & 0.10 & \\
73  & 5 38 41.2 & -69 08 52.1 & 13.6 & 0.02 & 13.0 & 0.03 & 12.8 & 0.03 & 12.6 & 0.07 & \\
74  & 5 38 41.2 & -69 02 58.2 & 13.8 & 0.04 & 13.7 & 0.06 & 13.6 & 0.06 & 13.0 & 0.08 & e, fg \\
75  & 5 38 41.3 & -69 08 43.4 & 16.3 & 0.14 & 15.4 & 0.14 & 14.7 & 0.14 & 13.6 & 0.17 & \\
76  & 5 38 41.3 & -69 05 32.3 & 13.2 & 0.05 & 13.1 & 0.08 & 13.1 & 0.09 & 13.8 & 0.13 & Parker 150 \\
77  & 5 38 41.4 & -69 03 54.0 & 15.7 & 0.08 & 13.3 & 0.02 & 11.5 & 0.02 & 08.8 & 0.04 & Brandner 16a, e \\
78  & 5 38 41.6 & -69 05 13.9 & 11.3 & 0.05 & 11.1 & 0.06 & 10.8 & 0.03 & 10.4 & 0.05 & e \\
79  & 5 38 42.1 & -69 05 55.3 & 11.9 & 0.18 & 11.9 & 0.22 & 11.6 & 0.13 & 11.5 & 0.07 & fg \\
80  & 5 38 42.2 & -69 08 04.2 & 13.8 & 0.02 & 13.0 & 0.02 & 12.9 & 0.02 & 13.0 & 0.09 & \\
81  & 5 38 42.2 & -69 06 14.4 & 11.6 & 0.08 & 11.6 & 0.13 & 11.0 & 0.06 & 11.0 & 0.05 & \\
82  & 5 38 42.2 & -69 08 32.3 & 13.6 & 0.04 & 13.3 & 0.06 & 13.1 & 0.04 & 12.8 & 0.07 & e \\
83  & 5 38 42.4 & -69 04 58.1 & 11.5 & 0.02 & 11.4 & 0.02 & 11.3 & 0.02 & 11.4 & 0.05 & fg \\
84  & 5 38 42.4 & -69 06 02.9 & 09.4 & 0.06 & 09.4 & 0.08 & 08.9 & 0.05 & 08.9 & 0.05 & R136\\
85  & 5 38 42.6 & -69 01 02.3 & 12.8 & 0.02 & 12.7 & 0.04 & 12.7 & 0.03 & 12.2 & 0.08 & e, fg \\
86  & 5 38 42.7 & -69 05 42.4 & 11.1 & 0.05 & 11.1 & 0.06 & 11.0 & 0.03 & 10.8 & 0.04 & fg \\
87  & 5 38 42.8 & -69 09 28.8 & 12.4 & 0.02 & 11.1 & 0.02 & 10.3 & 0.02 & 09.6 & 0.04 & \\
88  & 5 38 43.2 & -69 06 14.8 & 12.4 & 0.09 & 12.5 & 0.13 & 12.1 & 0.08 & 12.0 & 0.07 & e, fg \\
89  & 5 38 44.0 & -69 06 59.4 & 14.0 & 0.13 & 13.7 & 0.17 & 13.0 & 0.13 & 11.8 & 0.06 & e \\
90  & 5 38 44.2 & -69 05 46.7 & 12.4 & 0.10 & 12.6 & 0.19 & 12.3 & 0.10 & 12.1 & 0.06 & e, fg \\
91  & 5 38 44.3 & -69 06 05.8 & 12.1 & 0.02 & 11.8 & 0.05 & 11.5 & 0.04 & 11.1 & 0.05 & e \\
92  & 5 38 45.0 & -69 08 06.7 & 13.3 & 0.03 & 13.1 & 0.05 & 12.9 & 0.05 & 12.4 & 0.06 & e \\
93  & 5 38 45.1 & -69 04 46.9 & 14.2 & 0.12 & 14.0 & 0.16 & 13.5 & 0.17 & 12.4 & 0.09 & e \\
94  & 5 38 45.1 & -69 05 08.5 & 12.9 & 0.14 & 12.7 & 0.17 & 11.7 & 0.10 & 10.2 & 0.07 & Rubio 101, e \\
95  & 5 38 45.6 & -69 05 47.8 & 11.8 & 0.05 & 11.7 & 0.07 & 11.6 & 0.04 & 11.4 & 0.05 & e, fg \\
96  & 5 38 45.7 & -69 06 22.3 & 11.6 & 0.04 & 11.5 & 0.03 & 11.4 & 0.03 & 11.2 & 0.05 & fg \\
97  & 5 38 45.9 & -69 02 43.4 & 13.4 & 0.04 & 12.5 & 0.04 & 12.3 & 0.04 & 12.2 & 0.06 & \\
98  & 5 38 46.1 & -69 05 20.4 & 14.2 & 0.13 & 14.4 & 0.16 & 13.2 & 0.12 & 11.2 & 0.05 & Brandner 11b, e \\
99  & 5 38 46.1 & -69 04 55.2 & 15.4 & 0.35 & 14.5 & 0.36 & 13.8 & 0.24 & 11.2 & 0.09 & e \\
100 & 5 38 46.3 & -69 03 13.3 & 14.0 & 0.03 & 13.0 & 0.04 & 12.7 & 0.03 & 12.4 & 0.06 & \\
101 & 5 38 46.8 & -69 05 05.3 & 13.0 & null & 13.7 & 0.24 & 12.3 & 0.10 & 10.2 & 0.04 & e \\
102 & 5 38 47.1 & -69 05 01.7 & 12.9 & null & 13.3 & 0.20 & 11.4 & 0.08 & 08.8 & 0.04 & Rubio 126, Brandner 12d, e \\
103 & 5 38 47.6 & -69 08 48.8 & 13.0 & 0.02 & 11.9 & 0.03 & 11.5 & 0.03 & 11.4 & 0.05 & \\
104 & 5 38 48.1 & -69 04 11.6 & 15.2 & 0.17 & 13.1 & null & 12.0 & null & 10.2 & 0.05 & e \\
105 & 5 38 48.3 & -69 04 44.4 & 13.5 & 0.11 & 12.8 & 0.11 & 12.1 & 0.08 & 11.1 & 0.07 & Rubio 138, Brandner 15b, e \\
106 & 5 38 48.5 & -69 05 32.6 & 09.2 & 0.02 & 08.3 & 0.04 & 07.9 & 0.03 & 07.6 & 0.04 & \\
107 & 5 38 48.7 & -69 01 38.6 & 12.9 & 0.02 & 11.4 & 0.02 & 10.4 & 0.02 & 09.3 & 0.04 & e \\
108 & 5 38 49.6 & -69 03 43.9 & 16.7 & 0.18 & 15.8 & 0.19 & 15.2 & 0.19 & 13.4 & 0.13 & e \\
109 & 5 38 49.8 & -69 06 42.8 & 12.5 & 0.05 & 12.4 & 0.09 & 12.1 & 0.07 & 11.2 & 0.07 & e \\
110 & 5 38 50.0 & -69 03 38.2 & 14.0 & 0.03 & 13.4 & 0.04 & 12.9 & 0.03 & 12.2 & 0.07 & e \\
111 & 5 38 50.6 & -69 02 00.2 & 13.8 & 0.03 & 12.8 & 0.03 & 12.6 & 0.03 & 12.4 & 0.07 & \\
112 & 5 38 51.1 & -69 08 55.7 & 14.3 & 0.03 & 13.5 & 0.03 & 13.2 & 0.04 & 13.0 & 0.07 & \\
113 & 5 38 51.2 & -69 06 41.0 & 11.5 & 0.03 & 10.6 & 0.03 & 10.3 & 0.02 & 10.1 & 0.04 & \\
114 & 5 38 51.6 & -69 08 07.1 & 10.5 & 0.02 & 10.3 & 0.02 & 10.1 & 0.02 & 09.8 & 0.04 & e \\
115 & 5 38 52.6 & -69 11 24.4 & 12.4 & 0.02 & 11.3 & 0.02 & 11.0 & 0.02 & 10.6 & 0.05 & \\
116 & 5 38 52.7 & -69 04 37.6 & 16.3 & 0.20 & 14.9 & 0.14 & 13.6 & 0.06 & 11.2 & 0.06 & Rubio 169, e \\
117 & 5 38 52.9 & -69 03 21.2 & 15.6 & 0.08 & 14.3 & 0.07 & 13.8 & 0.07 & 13.6 & 0.13 & \\
118 & 5 38 53.4 & -69 02 00.6 & 10.9 & 0.02 & 10.8 & 0.02 & 10.6 & 0.02 & 10.1 & 0.04 & e, fg \\
119 & 5 38 53.6 & -69 04 58.8 & 13.6 & 0.18 & 13.5 & 0.25 & 12.1 & 0.12 & 10.9 & 0.08 & e \\
120 & 5 38 53.7 & -69 05 27.2 & 14.0 & 0.14 & 13.3 & 0.12 & 12.7 & 0.10 & 12.0 & 0.07 & e \\
121 & 5 38 53.8 & -69 03 32.0 & 14.2 & 0.04 & 13.3 & 0.03 & 13.1 & 0.04 & 13.0 & 0.09 & \\
122 & 5 38 53.9 & -69 09 31.3 & 15.6 & 0.09 & 14.9 & 0.11 & 14.1 & 0.08 & 11.7 & 0.05 & e \\
123 & 5 38 54.7 & -69 07 44.8 & 10.6 & 0.02 & 10.2 & 0.03 & 10.1 & 0.02 & 10.1 & 0.04 & \\
124 & 5 38 54.8 & -69 06 49.7 & 13.6 & 0.02 & 12.7 & 0.02 & 12.5 & 0.02 & 12.6 & 0.06 & \\
125 & 5 38 55.5 & -69 04 26.8 & 13.7 & 0.02 & 13.1 & 0.02 & 12.7 & 0.02 & 12.6 & 0.10 & \\
126 & 5 38 56.2 & -69 08 41.3 & 14.0 & 0.05 & 12.6 & null & 12.3 & null & 12.4 & 0.06 & \\
127 & 5 38 56.5 & -69 04 16.7 & 16.8 & null & 15.1 & null & 14.5 & 0.13 & 10.9 & 0.05 & e \\
128 & 5 38 57.1 & -69 06 05.4 & 11.2 & 0.02 & 11.1 & 0.03 & 10.9 & 0.02 & 10.6 & 0.04 & e \\
129 & 5 38 57.3 & -69 07 09.8 & 12.1 & 0.02 & 11.7 & 0.02 & 11.2 & 0.02 & 10.7 & 0.04 & e \\
130 & 5 38 57.7 & -69 10 39.7 & 13.6 & 0.03 & 12.5 & 0.03 & 12.1 & 0.03 & 11.4 & 0.07 & \\
131 & 5 38 58.3 & -69 04 21.4 & 15.3 & null & 15.1 & null & 14.3 & 0.14 & 12.0 & 0.06 & e \\
132 & 5 38 59.0 & -69 02 44.5 & 14.1 & 0.03 & 13.4 & 0.03 & 13.3 & 0.04 & 12.9 & 0.08 & \\
133 & 5 38 59.1 & -69 01 08.4 & 12.8 & 0.02 & 11.5 & 0.02 & 10.9 & 0.02 & 10.0 & 0.04 & \\
134 & 5 38 59.2 & -69 05 08.9 & 13.5 & null & 13.2 & null & 13.4 & 0.17 & 12.3 & 0.08 & e, fg \\
135 & 5 39 00.5 & -69 08 41.3 & 13.7 & 0.03 & 13.4 & 0.04 & 13.3 & 0.04 & 12.7 & 0.08 & e \\
136 & 5 39 01.0 & -69 07 58.8 & 13.7 & 0.04 & 12.8 & 0.05 & 12.5 & 0.04 & 12.5 & 0.05 & \\
137 & 5 39 01.9 & -69 02 34.1 & 14.6 & 0.04 & 13.4 & 0.04 & 12.9 & 0.03 & 12.4 & 0.07 & \\
138 & 5 39 03.8 & -69 03 46.4 & 14.9 & 0.02 & 14.3 & 0.04 & 13.6 & 0.04 & 13.6 & 0.16 & \\
139 & 5 39 04.8 & -69 04 09.8 & 12.3 & 0.03 & 12.2 & 0.04 & 12.1 & 0.04 & 11.7 & 0.06 & e, fg \\
140 & 5 39 05.3 & -69 04 16.0 & 15.5 & 0.11 & 15.7 & 0.22 & 15.1 & 0.20 & 13.3 & 0.15 & e \\
141 & 5 39 05.8 & -69 06 14.0 & 13.7 & 0.02 & 12.6 & 0.02 & 12.3 & 0.03 & 12.1 & 0.05 & \\
142 & 5 39 07.2 & -69 01 52.7 & 11.7 & 0.02 & 10.7 & 0.02 & 10.4 & 0.02 & 10.2 & 0.04 & \\
143 & 5 39 07.4 & -69 04 20.3 & 13.7 & 0.02 & 13.4 & 0.02 & 13.3 & 0.04 & 14.3 & 0.37 & \\
144 & 5 39 11.3 & -69 02 01.3 & 12.1 & 0.02 & 11.9 & 0.02 & 11.6 & 0.02 & 11.0 & 0.05 & e \\
145 & 5 39 11.4 & -69 08 25.1 & 14.1 & 0.02 & 12.4 & 0.02 & 11.2 & 0.02 & 10.2 & 0.04 & \\
146 & 5 39 12.5 & -69 04 08.8 & 13.2 & 0.02 & 12.8 & 0.03 & 12.7 & 0.03 & 12.6 & 0.07 & \\
147 & 5 39 12.5 & -69 02 09.6 & 12.8 & 0.02 & 12.6 & 0.02 & 12.5 & 0.03 & 12.3 & 0.07 & \\
148 & 5 39 14.3 & -69 05 25.1 & 15.1 & 0.04 & 14.2 & 0.05 & 13.7 & 0.05 & 13.6 & 0.14 & \\
149 & 5 39 14.6 & -69 05 03.5 & 12.5 & 0.02 & 11.5 & 0.02 & 11.3 & 0.02 & 11.3 & 0.05 & \\
150 & 5 39 16.3 & -69 05 28.7 & 14.0 & 0.03 & 13.0 & 0.03 & 12.7 & 0.04 & 12.5 & 0.07 & \\
151 & 5 39 16.3 & -69 07 34.3 & 14.4 & 0.02 & 13.2 & 0.02 & 12.8 & 0.02 & 12.8 & 0.06 & \\
152 & 5 39 18.1 & -69 05 16.4 & 14.2 & 0.04 & 13.4 & 0.04 & 13.3 & 0.05 & 13.1 & 0.10 & \\
153 & 5 39 18.2 & -69 08 48.5 & 13.8 & 0.03 & 13.6 & 0.04 & 13.6 & 0.05 & 13.3 & 0.12 & fg \\
154 & 5 39 19.1 & -69 04 22.4 & 14.5 & 0.03 & 13.5 & 0.04 & 13.3 & 0.04 & 13.4 & 0.12 & \\
155 & 5 39 19.8 & -69 10 10.9 & 13.1 & 0.02 & 12.0 & 0.02 & 11.4 & 0.02 & 10.9 & 0.05 & \\
156 & 5 39 20.2 & -69 04 17.0 & 15.2 & 0.07 & 14.2 & 0.07 & 13.9 & 0.07 & 13.5 & 0.18 & \\
157 & 5 39 20.2 & -69 06 26.3 & 12.0 & 0.02 & 11.9 & 0.02 & 11.8 & 0.03 & 11.7 & 0.05 & fg \\
158 & 5 39 25.8 & -69 11 35.9 & 09.0 & 0.02 & 08.4 & 0.05 & 08.3 & 0.02 & 08.2 & 0.04 & \\
159 & 5 39 26.0 & -69 06 29.2 & 14.8 & 0.05 & 14.4 & 0.08 & 14.0 & 0.09 & 14.3 & 0.27 & \\
160 & 5 39 28.2 & -69 05 50.6 & 11.4 & 0.02 & 10.7 & 0.03 & 10.6 & 0.02 & 10.6 & 0.04 & \\
161 & 5 39 28.9 & -69 06 56.9 & 14.6 & 0.03 & 13.7 & 0.03 & 13.4 & 0.05 & 13.3 & 0.09 & \\
162 & 5 39 29.3 & -69 05 18.6 & 15.1 & 0.07 & 14.3 & 0.05 & 14.0 & 0.08 & 13.7 & 0.15 & \\
163 & 5 39 29.7 & -69 03 42.1 & 13.1 & 0.02 & 12.2 & 0.02 & 11.9 & 0.02 & 11.6 & 0.06 & \\
164 & 5 39 32.0 & -69 04 40.8 & 13.1 & 0.02 & 12.0 & 0.03 & 11.6 & 0.02 & 11.2 & 0.05 & \\
165 & 5 39 32.1 & -69 05 43.8 & 13.8 & 0.03 & 12.7 & 0.03 & 12.3 & 0.03 & 12.1 & 0.05 & \\
166 & 5 39 32.9 & -69 03 40.7 & 14.4 & 0.03 & 13.3 & 0.03 & 13.0 & 0.04 & 12.6 & 0.07 & \\
167 & 5 39 33.6 & -69 08 55.0 & 13.5 & 0.02 & 12.7 & 0.02 & 12.4 & 0.02 & 12.4 & 0.05 & \\
168 & 5 39 34.4 & -69 10 44.8 & 13.8 & 0.02 & 12.8 & 0.03 & 12.5 & 0.02 & 12.1 & 0.08 & \\
169 & 5 39 35.5 & -69 04 39.0 & 13.5 & 0.02 & 12.9 & 0.03 & 12.7 & 0.03 & 12.8 & 0.08 & \\
170 & 5 39 35.8 & -69 04 08.0 & 13.6 & 0.02 & 12.5 & 0.03 & 12.1 & 0.02 & 11.9 & 0.05 & \\
171 & 5 39 36.1 & -69 05 15.4 & 14.0 & 0.03 & 13.1 & 0.03 & 12.8 & 0.03 & 12.9 & 0.08 & \\
172 & 5 39 36.5 & -69 08 48.8 & 13.5 & 0.02 & 12.7 & 0.02 & 12.5 & 0.02 & 12.7 & 0.07 & \\
173 & 5 39 37.8 & -69 05 01.0 & 12.4 & 0.02 & 11.3 & 0.03 & 11.0 & 0.02 & 10.8 & 0.05 & \\
174 & 5 39 37.9 & -69 11 46.3 & 10.9 & 0.02 & 10.3 & 0.04 & 10.1 & 0.06 & 09.8 & 0.05 & \\
175 & 5 39 38.5 & -69 09 00.4 & 14.2 & 0.03 & 13.6 & 0.02 & 12.3 & 0.03 & 10.5 & 0.06 & e \\
176 & 5 39 39.4 & -69 11 52.1 & 09.9 & 0.03 & 08.8 & 0.04 & 08.5 & 0.04 & 08.1 & 0.06 & \\
177 & 5 39 39.4 & -69 11 52.1 & 09.9 & 0.04 & 09.0 & 0.04 & 08.5 & 0.04 & 08.8 & 0.05 & \\
178 & 5 39 39.9 & -69 06 36.4 & 13.0 & 0.02 & 12.1 & 0.02 & 11.8 & 0.02 & 11.5 & 0.05 & \\
179 & 5 39 41.8 & -69 11 30.8 & 10.3 & 0.02 & 09.4 & 0.03 & 08.9 & 0.02 & 08.6 & 0.05 & \\
180 & 5 39 43.6 & -69 10 39.7 & 13.4 & 0.02 & 12.5 & 0.03 & 12.3 & 0.02 & 12.1 & 0.07 & \\
181 & 5 39 44.7 & -69 04 30.4 & 12.9 & 0.03 & 12.3 & 0.03 & 12.1 & 0.03 & 12.1 & 0.07 & \\
182 & 5 39 45.5 & -69 09 37.1 & 14.0 & 0.03 & 13.0 & 0.03 & 12.6 & 0.03 & 12.3 & 0.06 & \\
183 & 5 39 52.4 & -69 09 41.4 & 10.7 & 0.02 & 09.7 & 0.02 & 09.4 & 0.02 & 09.3 & 0.04 & \\
184 & 5 39 55.7 & -69 10 28.9 & 13.2 & 0.02 & 12.1 & 0.02 & 11.7 & 0.02 & 11.8 & 0.10 & \\
185 & 5 38 41.4 & -69 03 05.8 &  -   &  -   &  -   &  -   & 15.1 & 0.29 & 14.9 & 0.49 & \\
186 & 5 38 43.3 & -69 05 21.5 &  -   &  -   &  -   &  -   & 14.9 & 0.30 & 14.1 & 0.19 & \\
187 & 5 38 48.2 & -69 04 41.2 &  -   &  -   &  -   &  -   & 12.7 & 0.12 & 11.1 & 0.07 & Parker 1429, e \\
188 & 5 38 48.9 & -69 04 45.1 &  -   &  -   &  -   &  -   & 13.4 & 0.18 & 10.5 & 0.05 & e \\
189 & 5 38 50.5 & -69 05 02.0 &  -   &  -   &  -   &  -   & 12.8 & 0.14 & 11.7 & 0.09 & e \\
190 & 5 38 54.3 & -69 04 37.6 &  -   &  -   &  -   &  -   & 15.3 & 0.29 & 12.2 & 0.08 & e \\
191 & 5 39 35.1 & -69 06 28.8 &  -   &  -   &  -   &  -   & 15.6 & 0.37 & 14.1 & 0.21 & \\
192 & 5 38 15.3 & -69 06 54.0 &  -   &  -   &  -   &  -   &  -   &  -   & 12.6 & 0.08 & e \\
193 & 5 38 33.1 & -69 06 11.5 &  -   &  -   &  -   &  -   &  -   &  -   & 11.8 & 0.06 & Brandner 7a, e \\
194 & 5 38 34.1 & -69 04 52.3 &  -   &  -   &  -   &  -   &  -   &  -   & 10.5 & 0.04 & e \\
195 & 5 38 37.4 & -69 02 51.0 &  -   &  -   &  -   &  -   &  -   &  -   & 13.2 & 0.11 & e \\
196 & 5 38 40.5 & -69 05 57.5 &  -   &  -   &  -   &  -   &  -   &  -   & 12.0 & 0.06 & e \\
197 & 5 38 45.1 & -69 05 00.2 &  -   &  -   &  -   &  -   &  -   &  -   & 11.3 & 0.08 & Brandner 13b, e \\
198 & 5 38 45.3 & -69 04 41.9 &  -   &  -   &  -   &  -   &  -   &  -   & 12.0 & 0.06 & e \\
199 & 5 38 47.1 & -69 04 40.1 &  -   &  -   &  -   &  -   &  -   &  -   & 12.0 & 0.09 & e \\
200 & 5 38 48.8 & -69 04 40.4 &  -   &  -   &  -   &  -   &  -   &  -   & 11.1 & 0.07 & Brandner 14b, e \\
201 & 5 38 51.6 & -69 04 41.5 &  -   &  -   &  -   &  -   &  -   &  -   & 12.8 & 0.12 & e \\
202 & 5 38 54.5 & -69 05 00.2 &  -   &  -   &  -   &  -   &  -   &  -   & 10.9 & 0.09 & e \\
203 & 5 38 55.1 & -69 04 36.8 &  -   &  -   &  -   &  -   &  -   &  -   & 13.3 & 0.14 & \\
204 & 5 38 55.1 & -69 05 21.8 &  -   &  -   &  -   &  -   &  -   &  -   & 12.6 & 0.08 & e \\
205 & 5 38 55.9 & -69 05 38.0 &  -   &  -   &  -   &  -   &  -   &  -   & 11.5 & 0.06 & e \\
206 & 5 38 58.3 & -69 04 28.2 &  -   &  -   &  -   &  -   &  -   &  -   & 12.0 & 0.12 & e \\
207 & 5 38 59.9 & -69 04 19.6 &  -   &  -   &  -   &  -   &  -   &  -   & 12.5 & 0.12 & e \\
208 & 5 39 02.3 & -69 04 41.5 &  -   &  -   &  -   &  -   &  -   &  -   & 13.4 & 0.14 & \\
209 & 5 39 02.6 & -69 04 23.2 &  -   &  -   &  -   &  -   &  -   &  -   & 12.8 & 0.09 & e \\
210 & 5 39 08.1 & -69 04 05.2 &  -   &  -   &  -   &  -   &  -   &  -   & 13.1 & 0.10 & e \\
211 & 5 39 08.3 & -69 03 57.6 &  -   &  -   &  -   &  -   &  -   &  -   & 13.4 & 0.16 & \\
212 & 5 39 09.3 & -69 03 53.3 &  -   &  -   &  -   &  -   &  -   &  -   & 13.3 & 0.17 & \\
213 & 5 39 09.5 & -69 04 36.8 &  -   &  -   &  -   &  -   &  -   &  -   & 13.4 & 0.13 & \\
214 & 5 39 26.1 & -69 11 29.0 &  -   &  -   &  -   &  -   &  -   &  -   & 13.5 & 0.32 & \\
215 & 5 39 58.4 & -69 06 08.6 &  -   &  -   &  -   &  -   &  -   &  -   & 12.6 & 0.18 & e \\

\end{longtable}

\end{document}